\pdfoutput=1
\documentclass[journal,draftclsnofoot,onecolumn,12pt,twoside]{IEEEtranTCOM}
\normalsize

\ifCLASSINFOpdf
\else
\fi

\usepackage{graphicx}
\usepackage{amsmath}
\usepackage[mathscr]{eucal}

\usepackage{amssymb}
\usepackage{comment}
\usepackage{epsfig}
\usepackage{epstopdf}
\usepackage{algorithm}
\usepackage{algorithmic}
\usepackage{makecell} 

\usepackage{amsthm}
\usepackage{cite}
\usepackage{float}
\usepackage{color}
\usepackage{balance}

\newcommand{\Rabd}{\sum_{d \in \mathcal{U}_{\rm data}} R_{\rm AB}^d}
\newcommand{\Raedm}{\sum_{d\in \mathcal{U}_{\rm data}}R_{{\rm A}{\rm E}_m}^d}
\newcommand{\Sud}{\sum_{d \in \mathcal{U}_{\rm data}}}
\newcommand{\Pud}{\prod_{d \in \mathcal{U}_{\rm data}}}
\newcommand{\pAB}{P_{\rm AB}^{\rm OP}}
\newcommand{\psAB}{P_{\rm AB}^{\rm SOP}}
\newcommand{\psBA}{P_{\rm BA}^{\rm SOP}}

\begin{document}
%
\title{\large Novel Secret-Key-Assisted Schemes for Secure MISOME-OFDM Systems}
%
%
%

\author{\large Mohamed Marzban, ~\IEEEmembership{\large Student~Member,~IEEE}, Ahmed El Shafie, ~\IEEEmembership {\large Senior~Member,~IEEE}, 
Naofal Al-Dhahir, ~\IEEEmembership{\large Fellow,~IEEE} 
\vspace{-2\baselineskip}

\thanks{M. Marzban and N. Al-Dhahir are with the Electrical and Computer Engineering Department, the University of Texas at Dallas. A. El Shafie is with the R\&D department, Qualcomm Inc., USA. emails:(mohamed.marzban, aldhahir)@utdallas.edu, ahmed.salahelshafie@gmail.com}
}

\maketitle

\begin{abstract}
\vspace{-0.05 cm}

We propose a new secure transmission scheme for uplink multiple-input single-output (MISO) orthogonal-frequency multiplexing (OFDM) systems in the presence of multiple eavesdroppers. Our proposed scheme utilizes the sub-channels orthogonality of OFDM systems to simultaneously transmit data and secret key symbols. The base station, Bob, shares secret key symbols with the legitimate user, Alice, using wiretap coding over a portion of the sub-channels. Concurrently, Alice uses the accumulated secret keys in her secret-key queue to encrypt data symbols using a one time pad (OTP) cipher and transmits them to Bob over the remaining sub-channels. if Alice did not accumulate sufficient keys in her secret-key queue, she employs wiretap coding to secure her data transmissions. We propose fixed and dynamic sub-channel allocation schemes to divide the sub-channels between data and secret keys. We derive the secrecy outage probability (SOP) and the secure throughput for the proposed scheme. We quantify the system's security under practical non-Gaussian transmissions where discrete signal constellation points are transmitted by the legitimate source nodes. Numerical results validate our theoretical findings and quantify the impact of different system design parameters.
\end{abstract}
\vspace{-0.2cm}


\vspace{-0.2cm}
\begin{IEEEkeywords}
\vspace{-0.2cm}
PHY security, wiretap channel, OFDM, secret key sharing, OTP encryption
\end{IEEEkeywords}

\IEEEpeerreviewmaketitle

\vspace{-0.2cm}
\section{Introduction}
\IEEEPARstart{W}{ireless} channels are characterized by a broadcast nature which allows eavesdroppers to overhear exchanged messages between other users co-located in the same coverage area, introducing serious security challenges to the privacy of the users and the confidentiality of the transmitted data. Eavesdroppers usually conceal themselves as regular users and try to  eavesdrop the information transmitted by other users in the same network or coverage area. Hence, a system designer should expect possible eavesdropping attacks from some of the potential regular users. None of the users should be taken for granted as a non-hostile node. 


Information security is conventionally attained by deploying cryptography algorithms at the network layer of the protocol stack. Such algorithms have shown high resilience against eavesdropping attacks as they are hard to be broken in a feasible computational time. However, cryptography requires high storage and computational power resources at the legitimate communicating nodes. In addition, they are not information-theoretic secure since the computational power and knowledge of the network parameters are assumed to be restricted at the eavesdroppers. Recently, physical-layer (PHY) security has emerged as an additional layer of defense against eavesdroppers at the waveform level by exploiting the stochastic properties of the wireless channel. In \cite{wyner1975wire}, Wyner proved that information theoretic secure communications can be attained if the wiretap channel between the transmitter and the eavesdropper is statistically more degraded than the main channel between the transmitter and the legitimate receiver. The secrecy capacity, defined as the maximum data transmission rate with arbitrarily zero information exposure at the eavesdroppers, is used to quantify the system's security level.

Nowadays, orthogonal-frequency division multiplexing (OFDM) is being adopted by most wireless communications standards as it simplifies the channel equalization process by transforming the frequency-selective channel into a group of flat fading sub-channels. That is why many research works investigated the PHY security of OFDM systems, see e.g.  \cite{melki2019survey, wu2019physical, zhang2016energy, PlcWireless,wu2013practical,CCNC2,linearprecoder, scfdmaJournal, CCNC1, TVTcorrespondence, mukherjee2014principles, tvtjournal, japan}.  Some schemes relied on  designing artificial-noise (AN) precoding schemes to confuse the eavesdropping nodes and reduce their signal-plus-interference-to-noise ratios (SINRs). These schemes can be divided into spatial \cite{goel2008guaranteeing, wang_2017}, temporal ~\cite{scfdmaJournal, CCNC1, TVTcorrespondence}, or hybrid spatio-temporal \cite{tvtjournal} AN injection schemes. In all scenarios, the AN is injected in the null space of the legitimate receiver channel to eliminate its effect at the legitimate receiver. To achieve this goal, spatial AN injection schemes exploit the degrees of freedom provided by having multiple transmit antennas while the temporal AN schemes exploit the degrees of freedom due to the cyclic prefix (CP) insertion in the OFDM waveform. In \cite{scfdmaJournal}, the authors proposed a temporal AN injection scheme for multi-user systems and proved that the number of useful temporal AN streams is dependent on the channel memories of the transmitter-legitimate receiver and transmitter-eavesdropper links. In \cite{TVTcorrespondence}, the authors controlled the processing time delays in amplify-and-forward relay-based communications to increase the AN covariance matrix rank at the eavesdroppers. In \cite{tvtjournal}, the authors proposed a new hybrid spatio-temporal AN scheme to enhance the system's secrecy rate.


Other approaches for PHY security utilized the randomness and reciprocity properties of the wireless channel to generate identical secret key pairs at both the legitimate transmitter and receiver ends \cite{ref1xx,ref4xx,ref6xx,ref7xx,zhang2016efficient}. The transmitter encrypts its data symbols using these keys through the one time pad (OTP) cipher scheme \cite{ref15xx}. Different  properties of the received signals were investigated for secret key generation such as the received signal strength \cite{ref1xx}, phase differences \cite{ref4xx}, time delay (in wide-band transmission) \cite{ref6xx,ref7xx}, and channel state information (CSI) \cite{zhang2016efficient}. OTP encryption is mathematically unbreakable as long as the the number of secret key symbols is equal to the number of data symbols. However, the channel-based key generation rate is typically much lower than the data symbols rate which limits the secure transmission rate of these schemes.

In \cite{khalil2009delay}, the authors proposed the idea of using a secret-key queue in a single-user system. Different from the data queue at the transmitter, a secret-key queue is kept at both the legitimate transmitter and the receiver while being hidden from the eavesdropper. The main idea is to use a portion of the secrecy rate to send randomly-generated secret key symbols instead of sending data symbols. These stored key symbols can be used later to achieve secure communication between the source and its destination when the transmitter-destination link is not secured. Thus, even at times of \emph{zero} instantaneous secrecy rate, which is the case when the transmitter-eavesdropper channel has higher gain than that of the legitimate transmitter-receiver channel, data symbols can still be transmitted and secured from the eavesdropper. Based on this idea, Gungor \emph{et al.} \cite{gungor2013secrecy} showed that a long-term constant secrecy rate is achievable. The authors of \cite{gungor2013secrecy} addressed decoding delays but did
not consider the dynamics of the data arrival process at the transmitter. Mao \emph{et al.} \cite{mao2013achieving} assumed random data arrivals at the source node and illustrated through simulations that the use of a secret-key queue reduces the queuing delay for the data packets.

 Unlike \cite{khalil2009delay,gungor2013secrecy,mao2013achieving}, we propose a secret-key assisted secure scheme that utilizes the sub-channel orthogonality of OFDM systems, allowing the simultaneous transmissions of data and secret key symbols in two opposite directions. In addition, we consider non-Gaussian transmissions, where the data and keys are generated from discrete signal constellations, which is a practical hardware constraint for transmitters using a fixed modulation scheme. 
Our main contributions in this paper are summarized as follows
\begin{itemize}
	\item We propose a new secret-key-assisted transmission scheme where a base-station, Bob, securely transmits secret key packets to a legitimate source node, Alice, using a wiretap coding scheme. Those packets are accumulated in a secret-key queue at Alice and Bob. Alice then utilizes her secret keys to encrypt her data symbols through an OTP cipher scheme and transmits them to Bob. Our proposed secret-key-assisted scheme is designed based on the connection outage status of the Alice-Bob link.
	\item We propose two schemes to divide the OFDM sub-channels between Alice-Bob data transmissions and Bob-Alice secret key sharing. The first scheme optimizes a fixed sub-channel allocation and the target data/key rates based on the channel statistics. The second scheme is dynamic and optimizes the sub-channel allocation and the target data/key rates every coherence time. We show that the fixed sub-channel allocation scheme achieves close performance to the dynamic sub-channel allocation scheme using lower overhead and complexity. Moreover, we optimize the secret key packet length. We show the impact of all those system design parameters on the secure throughput.

    \item We derive the secrecy rate and the secrecy outage probability (SOP) expressions under gap approximations where non-Gaussian transmissions are adopted by the legitimate transmitting nodes (i.e., source node when it sends data packets and the base-station when it sends key packets).
    
    \item We analyze the Markov chain of the secret-key queue and show the impact of the system parameters on the queue dynamics and the amount of arrival and departure rates.
\end{itemize}

\emph{\underline{Notation:}}  Lower- and upper-case bold letters denote column vectors and matrices, respectively. 
$\mathbb{C}^{M \times N}$ denotes the set of all $M\times N$ complex matrices. $(\cdot)^\top$ and $(\cdot)^*$ denote the transpose and Hermitian (i.e., complex-conjugate transpose) operations, respectively, $\|.\|$ denotes the Euclidean norm of a vector.
$[\cdot]^+=\max\{0,\cdot\}$ returns the maximum between the argument and zero. The superscripts $(\cdot)^d$ and $(\cdot)^k$ respectively, refer to $d$-th data and $k$-th secret key sub-channels quantities. The subscripts $(\cdot)_{\rm m_1-m_2}$ denotes a quantity associated with $m_1 - m_2$ link where $m_1,m_2 \in \{{\rm Alice, Bob, \: } m{\rm -th \: Eve}\}$

\section{System model and proposed secret-key-assisted scheme}
\vspace{-0.05cm}
We consider an uplink OFDM-based wireless communication system where a legitimate user, Alice, transmits her data to a base-station (BS), Bob, through a set of orthogonal OFDM sub-channels   in the presence of $M$ other users who can be treated as potential non-colluding eavesdroppers, Eves. Let ${\rm A}$, ${\rm B}$, ${\rm E}_{m}$ denote Alice, Bob and the $m$-th Eve ($m\in\{1,2,\dots,M\}$), respectively. We consider a multiple-input single-output (MISO) two-way communications system where each transceiver is equipped with multiple transmit antennas and a single receive antenna. Let the number of transmit antennas at Alice and Bob be denoted by $N_{\rm A}$ and $N_{\rm B}$, respectively. 

Let $\mathbf{H}_{m_{\rm 1}-m_{\rm 2}}^{\rm time} \in \mathbb{C}^{N_{\rm m_{\rm 1}} \times L}$ denote the multi-path channel coefficients matrix from node $m_{\rm 1}$ to node $m_{\rm 2}$ where $m_{\rm 1}, m_{\rm 2} \in \{{\rm A,B,E}_{m}\}$, the $\nu$-th row represents the channel impulse response (CIR) with $L$ taps from transmit antenna $\nu$ to the receive antenna. Each CIR tap follows a complex Gaussian distribution. We assume a quasi-static fading channel model in which the channel experiences slowly-varying fading and the entire codeword is experiencing a single fading channel realization. The OFDM system decouples a frequency-selective channel into a set of orthogonal flat fading sub-channels. The transmitter performs an inverse discrete Fourier transform (IDFT) of size $N$ and inserts a CP of size $N_{\rm cp} \geq L$. At the receivers side, the CP is removed and a DFT operation is performed to obtain the frequency domain signal. The corresponding frequency-domain channels are divided into $N$ orthogonal sub-channels. Let the channel frequency response between a transmitter, $m_{\rm 1}$, and a receiver, $m_{\rm 2}$, over a sub-channel, $k$, be denoted by $\mathbf{h}_{\rm m_{\rm 1}-m_{\rm 2}}^k \in \mathbb{C}^{1 \times N_{\rm m_{\rm 1}}}$ where $m_{\rm 1}, m_{\rm 2} \in \{{\rm A,B,E}_{m}\}$. Note that $\mathbf{h}^k_{\rm m_{\rm 1}-m_{\rm 2}}$ is not the same as $\mathbf{h}^k_{\rm m_{\rm 2}-m_{\rm 1}}$.

We assume that Alice and Eves are potential users in the network communicating with the BS, Bob, who can estimate his links' channels with both Alice and Eves. This assumption is reasonable since Alice and Eves are network users, feeding back their channel response estimates to Bob. In contrast, the channel state information (CSI) of the Alice-Eves links are completely unknown at both Alice and Bob. Hence, Alice cannot design her precoder based on any knowledge of the Alice-Eves links' channels. It is noteworthy that Alice could design an AN signal to hurt Eves using the knowledge of Alice-Bob link where the AN lies in a subspace orthogonal to the subspace spanned by the Alice-Bob channel vector \cite{tvtjournal,scfdmaJournal}. However, we skip this design in this paper since it is not going to change our scheme's design or add any insight to the problem under investigation. 

\begin{figure}
    \centering
\normalcolor
  \includegraphics[width=0.8\columnwidth]{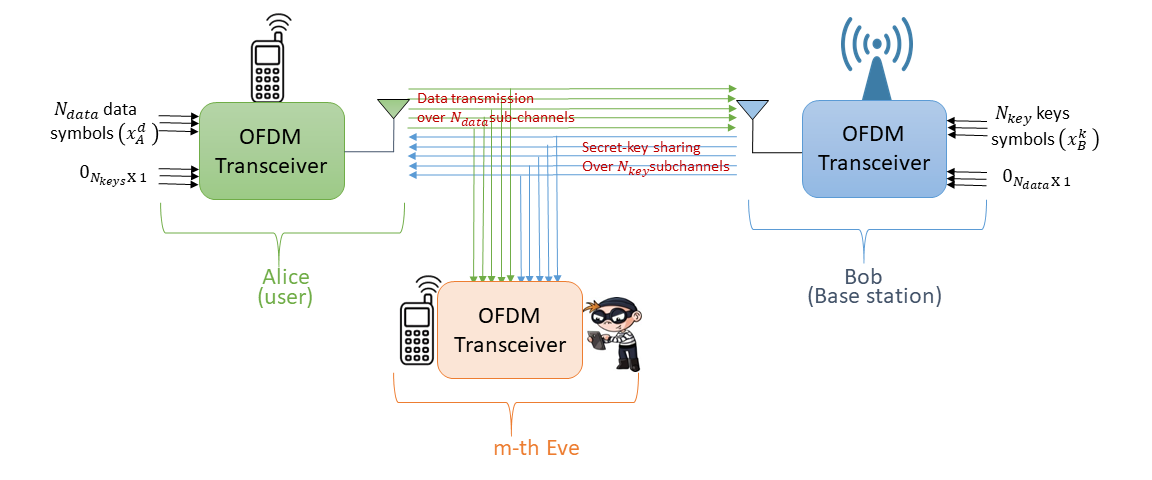}
    \vspace{-0.5 cm}
  \caption{System model showing Alice sharing data to Bob using $N_{\rm data}$ sub-channels and Bob sharing secret keys to Alice using $N_{\rm keys}$ sub-channels in the presence of $M$ eavesdroppers who can eavesdrop on both the data and the secret-key transmissions.}\label{figSharedOFDM}
  \vspace{-0.8 cm}
\end{figure}

The key idea of our proposed scheme is to exploit the orthogonal structure of the OFDM sub-channels to simultaneously transmit data and secret keys over different OFDM sub-channels. We assume two sets of OFDM sub-channels. The set of data OFDM sub-channels, denoted by ${\mathcal{U}_{\rm data}}$, and has a cardinality of $N_{\rm data}$, and the set of the secret key OFDM sub-channels, denoted by ${\mathcal{U}_{\rm key}}$, and has a cardinality of $N_{\rm key}$. This imposes a constraint that $N_{\rm key}+N_{\rm data}=N$ where $N$ is the total number of OFDM sub-channels\footnote{In practice, some sub-channels are not used (to create guard bands) and others are dedicated for control functions such as channel estimation and synchronization. In this case, $N$ will represent the number of active sub-channels dedicated for data transmissions that does not include the control sub-channels}. Alice uses the $N_{\rm data}$ sub-channels to transmit data to Bob and, simultaneously, Bob uses the remaining $N_{\rm key}=N-N_{\rm data}$ OFDM sub-channels to transmit secret-key symbols to Alice as illustrated in Fig. \ref{figSharedOFDM}. Since the OFDM sub-channels are orthogonal, there is no interference between the two transmissions. 

Let $d \in \mathcal{U}_{\rm data}$ and $k \in \mathcal{U}_{\rm key}$ represent a sub-channel assigned for data transmission and secret key sharing, respectively. Let Alice's transmitted data symbol over the $d$-th sub-channel be denoted by $x_{\rm A}^d$ and the secret-key symbol shared by Bob over the $k$-th sub-channel be denoted by $x_{\rm B}^k$. Hence, the received data signal at Bob and the received secret-key symbol at Alice are given, respectively, by
{
\begin{equation}
y^d_{\rm B} = \mathbf{h}_{\rm AB}^{d \top} \: \mathbf{p}_{\rm A}^d x_{\rm A}^d + n_{\rm B}^d, \;\;\;\; 
y^k_{\rm A} = \mathbf{h}_{\rm BA}^{k \top} \: \mathbf{p}_{\rm B}^k x_{\rm B}^k + n_{\rm A}^k 
\end{equation}
}
where { $\mathbf{h}_{\rm AB}^{d} \in \mathbb{C}^{N_{\rm A} \times 1}$ and $\mathbf{h}_{\rm BA}^{k} \in \mathbb{C}^{N_{\rm B} \times 1}$} denote the Alice-Bob and Bob-Alice channel frequency response vectors, at the $d$-th and $k$-th sub-channels, respectively. Let $\mathbf{p}_{\rm A}^d \in \mathbb{C}^{\rm N_{\rm A} \times 1} $ and $\mathbf{p}_{\rm B}^k \in \mathbb{C}^{\rm N_{\rm B} \times 1} $ denote the $k$-th and $d$-th sub-channel precoding vectors at Alice and Bob, respectively. Finally, let $n_{\rm A}^d$ and $n_{\rm B}^k$  denote complex additive white Gaussian noise (AWGN) samples at Alice and Bob, respectively. 

We assume that the target secret-key rate and the target data rate are fixed, respectively, at $\mathcal{R}_{\rm key}$ and $\mathcal{R}_{\rm data}$ bits/channel use. Generally speaking, $\mathcal{R}_{\rm data}$ is not necessarily equal to $\mathcal{R}_{\rm key}$. To simplify the relationship between both quantities, we assume that $\mathcal{R}_{\rm key}$ is divisible by $\mathcal{R}_{\rm data}$. More specifically, $\mathcal{R}_{\rm data}/\mathcal{R}_{\rm key}= K $ where $K \in \{1,2,\dots\}$ is an integer. Note that when the direct link from Alice to Bob is in a connection outage, the link from Bob to Alice is not necessarily in a connection outage since the channels and precoders of both links are different. In addition, $\mathcal{R}_{\rm key}$ is typically designed to be less than $\mathcal{R}_{\rm data}$ which constrains the secret-key packet length to be relatively smaller than the data packet length.

\begin{figure}
    \centering
\normalcolor
  \includegraphics[width=0.5\columnwidth]{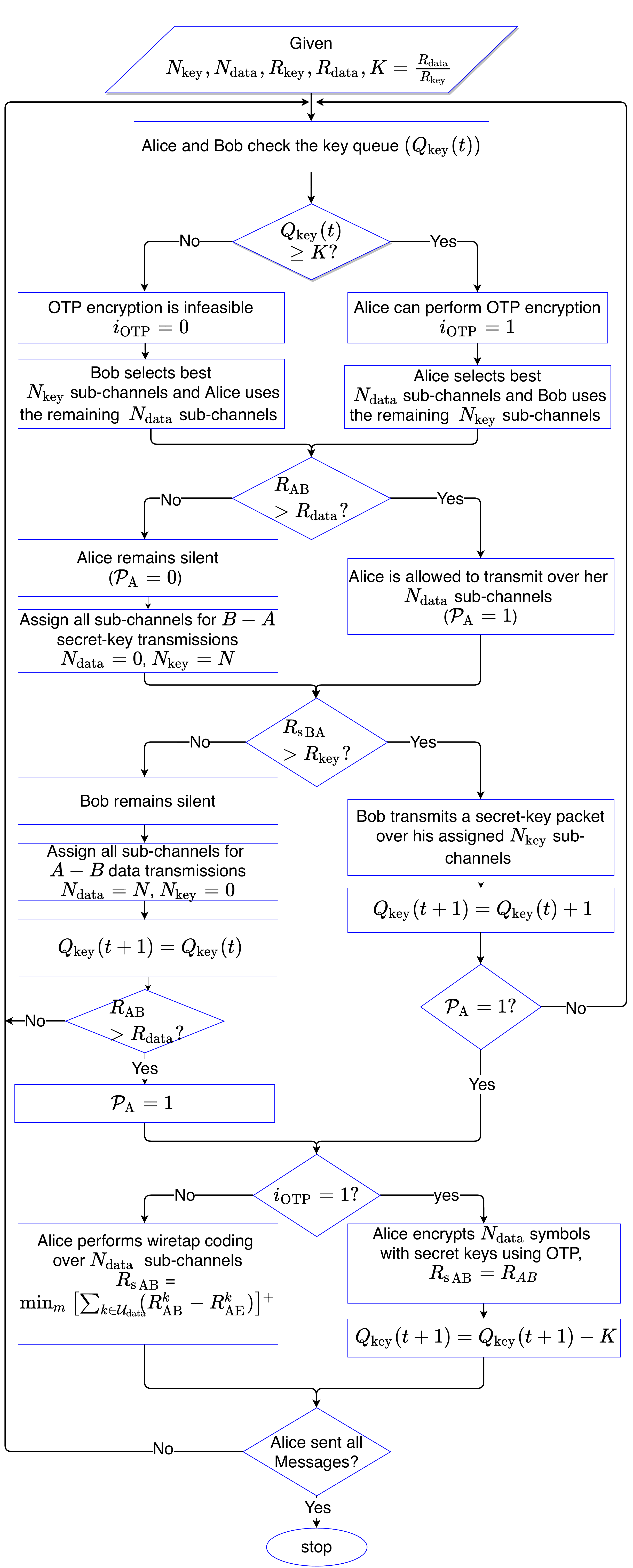}
  \caption{Flow chart of our proposed scheme where $\mathcal{P}_{\rm A} = 1$ implies $A-B$ link is not in connection outage and thus Alice is allowed to transmit data and $i_{\rm OTP} = 1$ implies Alice accumulated sufficient keys for OTP encryption. }\label{fig:flowChart}
\end{figure}

The flow chart of the proposed scheme is depicted in Fig. \ref{fig:flowChart}. Let $Q_{\rm key} (t)$ denote Alice's secret-key queue state at time $t$ measured by the number of received secret-key packets from Bob. If the number of accumulated secret-key packets is greater than a single data packet size (i.e., $Q_{\rm key} (t) \geq K$), Alice selects the $N_{\rm data}$ sub-channels having highest channel gains of the Alice-Bob link. 
 Note that the best $N_{\rm data}$ sub-channels of the Alice-Bob link are not necessarily the best sub-channels of the Bob-Alice link. Then, Alice utilizes $K$ secret-key packets in her secret-key queue to encrypt a single data packet using OTP encryption. In this case, the encrypted data symbols transmitted from Alice to Bob are perfectly secured and depend only on the Alice-Bob channel link irrespective of the Eve's received signals or the links' qualities. More precisely, the achieved Alice-Bob secrecy rate, ${R_{\rm s}}_{\rm AB}$, becomes equal to the Alice-Bob link rate, $R_{\rm AB}$, as follows
\begin{equation}
    {R_{\rm s}}_{\rm AB} = \sum_{d \in \mathcal{U}_{\rm data}} R_{\rm AB}^{d}
\end{equation}
If the direct link is not in a connection outage (i.e. $ R_{\rm AB} > \mathcal{R}_{\rm data}$), the data transmission is successful and the secret-key queue moves to state $Q_{\rm key}(t) - K$, i.e., $K$ packets leave the queue's head as follows
\begin{equation}
    Q_{\rm key} (t+1) = Q_{\rm key} (t) - K
\end{equation}
For secret-key sharing, Bob utilizes knowledge of his channels with Alice and Eves to adjust the redundancy rate of the wiretap codes to send perfectly-secured secret-key symbols to Alice over the remaining $N_{\rm key}$ sub-channels. The details of the wiretap code design are discussed in Appendix \ref{wiretapCoding}. Note that those $N_{\rm key}$ sub-channels are assigned to key sharing only if the $\rm B-A$ achieved secrecy key rate, ${R_{\rm s}}_{\rm BA}$, is higher than the target secret-key rate, $\mathcal{R}_{\rm key}$. Otherwise, all sub-channels are assigned to data transmission (i.e. $N_{\rm data} = N$ and $N_{\rm key} = 0$).

When the number of accumulated secret-key packets at Alice/Bob is less than a single data packet size (i.e., $Q_{\rm key} (t) < K$), OTP encryption is not possible, since OTP requires an equal number of secret-key and data symbols. Hence, data transmissions are secured using wiretap coding, which is considered secure as long as the achieved secrecy data rate, ${R_{\rm s}}_{\rm AB}$, is higher than the target data rate, $\mathcal{R}_{\rm data}$. Otherwise, the system is considered to be in a secrecy outage, as discussed in Appendix \ref{wiretapCoding}. In case of non-sufficient secret key packets in the secret-key queue ($Q_{\rm key} (t) < K$), Bob is given the priority in selecting the $N_{\rm key}$ sub-channels used for secret-key transmissions. Bob selects the $N_{\rm key}$ sub-channels that maximize the difference between the ${\rm B}-{\rm A}$ link rate, ${R}_{\rm BA}^{ k}$,  and the maximum link rate among the Bob-Eves links, $\max_{m} R_{{\rm BE}_m}^{ k}$, as follows
\begin{equation} \label{BobSelects1}
\small
   \begin{split}
   \mathcal{U}_{\rm key}(i) &= {\arg} { \max_{k}} \; \left\{ R_{\rm BA}^{ k} - \max_{m} R_{{\rm BE}_m}^{ k}\right\}, \forall i \in \{1,2,.., N_{\rm key}\}, k \neq \left\{\mathcal{U}_{\rm key}(1),.., \mathcal{U}_{\rm key}(i-1)\right\}
   \end{split}
\end{equation}
The flow chart of our proposed scheme is depicted in Fig. \ref{fig:flowChart} and the list of key variables in this paper is summarized in Table \ref{Table:variables}.

\begin{table*}
\fontsize{8}{5}\selectfont
\caption{List of Key Variables}
\vspace{-0.5 cm}
\label{Table:variables}
\centering
    \begin{tabular} { | p{1.3 cm} p{6.3 cm} | p{1.3 cm} p{6.3 cm} |}
    \hline
    symbol &  Description &  Symbol & Description \\ \hline\hline
    $N$ & total number of sub-channels  &  $N_{\rm cp}$ & CP length \\
    $N_{\rm data}$ & number of data sub-channels  &  $N_{\rm key}$ & number of secret keys sub-channels \\
    ${\mathcal{U}_{\rm data}}$ & set of data sub-channels  &  ${\mathcal{U}_{\rm key}}$ & set of key sub-channels \\
    $N_{\rm A}$ & number of transmit antennas at Alice  &  $N_{\rm B}$ & number of transmit antennas at Bob \\
    $M$ & number of eavesdroppers  &  $L$ & number of CIR taps \\
    $\mathbf{H}_{m_{\rm 1}-m_{\rm 2}}^{\rm time}$ & multipath CIR of size $ N_{\rm m_{\rm 1}} \times L$ from node $m_{\rm 1}$ to node $m_{\rm 2}$  &  $\mathbf{h}_{m_{\rm 1}-m_{\rm 2}}^{k} $ & frequency domain channel  from node $m_{\rm 1}$ to node $m_{\rm 2}$ over sub-channel $k$\\
    $\mathbf{p}_{\rm A}^d$ & $d$-th sub-channel precoding vector at Alice of length $N_{\rm A}$ &  $\mathbf{p}_{\rm B}^k$ & $k$-th sub-channel precoding vector at Bob of length $N_{\rm B}$   \\
    $\mathcal{R}_{\rm data}$ & target data rate &  $\mathcal{R}_{\rm key}$ & target key rate \\
    $Q_{\rm key}(t)$ & secret-key queue state at time $t$ &  $Q_{\rm max}$ & size of the secret-key queue\\
    $K$ & ratio between $\mathcal{R}_{\rm data}$ and $\mathcal{R}_{\rm key}$ which is equal to the ratio between the data and key packet sizes  & $\lambda$ &  mean arrival rate of the secret-key queue \\
    ${R}_{\rm AB}$ & achieved data rate for ${\rm A}-{\rm B}$ link  &   ${R}_{\rm BA}$ & achieved key-sharing rate for ${\rm B}-{\rm A}$ link \\
    ${R_{\rm s}}_{\rm AB}$ & achieved secrecy data rate for ${\rm A}-{\rm B}$ link   &  ${R_{\rm s}}_{\rm BA}$ & achieved secrecy key rate for ${\rm B}-{\rm A}$ link \\
    $\gamma^d_{\rm A}$ & $d$-th sub-channel received SNR for ${\rm A}-{\rm B}$ transmission  & $\gamma^k_{\rm B}$ & $k$-th sub-channel received SNR for ${\rm B}-{\rm A}$ transmission \\
    $\Gamma_{\rm AB}$ & SNR gap for ${\rm A}-{\rm B}$ transmissions  &  $\Gamma_{\rm BA}$ & SNR gap for ${\rm B}-{\rm A}$ transmission \\
    $\mathcal{T}_{\rm s}$ & achieved secure throughput  &   $\pAB$ & connection outage probability for ${\rm A}-{\rm B}$ data transmission  \\
    $\psAB$ & SOP for ${\rm A}-{\rm B}$ data transmissions  &  $\psBA$ & SOP for ${\rm B}-{\rm A}$ secret key sharing \\
     \hline\hline
    \end{tabular}
    \vspace{-0.5 cm}
\end{table*}

\section{Secrecy outage probabilities (SOP)}

The flow chart in Fig. \ref{fig:flowChart} shows that in some scenarios Alice employs OTP encryption over its assigned sub-channels while in others, wiretap coding is adopted. However, Bob always adopts a wiretap coding scheme to share his generated secret keys with Alice. In this section, we derive the SOP for these three scenarios: OTP encryption for Alice transmissions to Bob, wiretap coding for Alice transmissions to Bob, and wiretap coding for Bob transmissions to Alice.

\subsection{Alice-Bob link SOP using OTP encryption}
Alice's data queue is assumed to be backlogged with data packets. Considering a channel bandwidth of $W$ and a time slot duration $T$, each packet has ${B}_{\rm data}=\mathcal{R}_{\rm data} W T$ bits where the product $W T$ represents the total number of channel uses and $\mathcal{R}_{\rm data}$ is the number of bits per channel use. The achievable data rate of the Alice-Bob link over the $d$-th OFDM sub-channel is given by
\begin{equation}\label{gamma_2}
\begin{split}
R_{\rm AB}^d = \frac{1}{N+N_{\rm cp}} \log_2 \left(1+\gamma^d_{\rm A} \frac{\mathbf{h}_{\rm AB}^{d \top} \mathbf{p}_{\rm A}^d \left(\mathbf{h}_{\rm AB}^{d \top}\mathbf{p}_{\rm A}^d\right)^* }{\Gamma_{\rm AB}}\right)
\end{split}
\end{equation}
where $\gamma^d_{\rm A}$ is the signal-to-noise ratio (SNR) at the $d$-th sub-channel and $\Gamma_{\rm AB}$ is the SNR gap for the ${\rm A}-{\rm B}$ link. The SNR gap accounts for the mismatch between the theoretical capacity based on error free transmissions of Gaussian signals and achievable rates using practical coded quadrature amplitude modulation (QAM) schemes. The SNR gap is given by\cite{CioffiNotes_2}
\begin{align}\label{gamma_1}
\Gamma_{\rm AB} =\frac{\gamma_m}{3 \gamma_c} \left(Q^{-1}\left(\frac{P_{\rm AB,e}}{4}\right)\right)^2
\end{align}
where $Q(x)=\frac{1}{\sqrt{2 \pi}}\int_{x}^{\infty} \text{exp}(-\frac{u^2}{2}) du$ is the well-known $Q$-function, $\gamma_c$ is the coding gain of any applied code, $\gamma_m$ is the design margin to ensure adequate performance in the presence of unforeseen system impairments, and $P_{\rm AB,e}$ is the target error rate. The value of the SNR gap, $\Gamma_{\rm AB}$, defines the required system quality-of-service (QoS). Therefore, we can define different gap values for each group to enable different users priority levels.

 Alice designs her $d$-th sub-channel precoder, $\mathbf{p}_{\rm A}^d$,  based on the CSI of the ${\rm A}-{\rm B}$ link as follows, $\mathbf{p}_{\rm A}^d=\frac{(\mathbf{h}_{\rm AB}^d)^*}{|\mathbf{h}_{\rm AB}^d|}$. The achievable ${\rm A}-{\rm B}$ link rate over the $d$-th OFDM sub-channel is given by
\begin{equation}\label{gamma_xx}
\begin{split}
R_{\rm AB}^d =  \frac{1}{N+N_{\rm cp}} \log_2 \left(1+\gamma^d_{\rm A} \frac{\|\mathbf{h}_{\rm AB}^d\|^2 }{\Gamma_{\rm AB}}\right)
\end{split}
\end{equation}
When the number of Alice's transmit antennas is large, $\|\mathbf{h}_{AB}^d\|^2$ can be approximated by the expectation of its elements (i.e. $\|\mathbf{h}_{AB}^d\|^2\approx N_{\rm A}$). Consequently,
\begin{equation}\label{rateapprox1}
\begin{split}
R_{\rm AB}^d \approx  \frac{1}{N+N_{\rm cp}} \log_2 \left(1+\gamma^d_{\rm A} \frac{N_{\rm A}}{\Gamma_{\rm AB}}\right)
\end{split}
\end{equation}
In OTP encryption, the transmitted symbols are perfectly secured from Eves and the secrecy rate becomes equal to the data rate (i.e. ${R^d_{\rm s}}_{\rm AB} =  R_{\rm AB}^{d}$). Hence, under OTP encryption, the secrecy outage probability, $P_{\rm AB}^{\rm SOP}$, is equal to the connection outage probability, $P_{\rm AB}^{\rm OP}$, as follows 
\begin{equation}
\label{connOutage}
P_{\rm AB}^{\rm OP} = \Pr \left \{ \sum_{d \in \mathcal{U}_{\rm data}} R_{\rm AB}^d \leq \mathcal{R}_{\rm data} \right\}
\end{equation}
To ensure that Alice's transmissions are not in connection outage, the following condition should be satisfied.
\begin{equation}\label{rateapproxxx}
\begin{split}
\sum_{d \in \mathcal{U}_{\rm data}} R_{\rm AB}^d \approx \frac{N_{\rm data}}{N+N_{\rm cp}}\log_2 \left(1+\gamma^d_{\rm A} \frac{N_{\rm A}}{\Gamma_{\rm AB}}\right)\ge \mathcal{R}_{\rm data}
\end{split}
\end{equation}
which implies that
\begin{equation}\label{rateapproxxx2} 
\begin{split}
 \gamma^d_{\rm A} \frac{N_{\rm A}}{\Gamma_{\rm AB}}\ge 2^{\mathcal{R}_{\rm data} \frac{N+N_{\rm cp}}{N_{\rm data}}}-1
\end{split}
\end{equation}
Alice can satisfy this constraint by controlling the number of data OFDM sub-channels, $N_{\rm data}$, the number of transmit antennas, $N_{\rm A}$, the transmit power, $\gamma^k_{\rm A}$, and the SNR gap, $\Gamma_{\rm AB}$, which is a function of the required error rate. Under this condition, the probability of successful (reliable) transmission is almost one. In other words, the connection outage, which is equal to the SOP in this case, is almost zero. 
\subsection{Alice-Bob link SOP using wiretap coding}
If Alice does not have sufficient keys to encrypt one packet of her data symbols using OTP encryption, she employs wiretap coding over the $N_{\rm data}$ sub-channels. In this case, the instantaneous secrecy rate of the ${\rm A}-{\rm B}$ link is given by \cite{khisti_2010}
\begin{equation}\label{RAB1}
\sum_{d \in \mathcal{U}_{\rm data}}   {R_{\rm s}}_{\rm AB}^d =  \frac{1}{N+N_{\rm cp}} \min_m \left[\sum_{d \in \mathcal{U}_{\rm data}}\left(R_{\rm AB}^d- R_{{\rm A}{\rm E}_m}^d\right)\right]^+
\end{equation}
where $R_{{\rm A}{\rm E}_m}^d$ is the achievable rate at the $m$-th Eve over the $d$-th OFDM sub-channel. Alice does not know ${\rm A}-E_m$ channel links and cannot compute $ {R_{\rm s}}_{\rm AB}^d$. Hence, perfect secrecy cannot be attained and a system designer should expect an SOP which can be given by 
\begin{equation}\label{NSOP1}
\small
\begin{split}
P_{\rm AB}^{\rm SOP} &= 1-  \Pr\{ {R_{\rm s}}_{\rm AB} \geq \mathcal{R}_{\rm data} \} = \Pr \bigg\{ \Big( \Rabd- \max_m  \Raedm \Big)\ < \mathcal{R}_{\rm data} \! \bigg\}
\end{split}
\end{equation}
\normalsize
The achievable rate of the Alice to $m$-th Eve link, dented by ${\rm A}-{\rm E}_m$, can be given by\footnote{We emphasize that the Alice-Eve's achievable data rate is arbitrarily small whenever Alice-Bob's link rate is higher than Alice-Eve's link rate as Alice employs wiretap coding. When the secrecy target rate is lower than the instantaneous secrecy rate, the information leakage at Eve is zero.}
\begin{equation}\label{AE1}
\small
\begin{split}
& \Sud \!\!\!  R_{{\rm A}{\rm E}_m}^d = \frac{\Sud \!\!\! \log_2 \left(1+\gamma^d_{\rm A} \mathbf{h}_{{\rm A}{\rm E}_m}^{d \top} \mathbf{p}_{\rm A}^d \left(\mathbf{h}_{{\rm A}{\rm E}_m}^{d \top}\mathbf{p}_{\rm A}^d\right)^* \right)}{N+N_{\rm cp}}   = \frac{\log_2   \Pud \left(1+\gamma^d_{\rm A} \mathbf{h}_{{\rm A}{\rm E}_m}^{d \top}  \mathbf{p}_{\rm A}^d \left(\mathbf{h}_{{\rm A}{\rm E}_m}^{d \top}\mathbf{p}_{\rm A}^d\right)^* \right)}{N+N_{\rm cp}} 
\end{split}
\end{equation} 
\normalsize
Substituting with \eqref{rateapprox1} and \eqref{AE1}  into \eqref{NSOP1}, we get
\begin{equation}
\begin{split}
     & P_{\rm  AB}^{\rm SOP}  = \Pr\bigg\{\frac{1}{N+N_{\rm cp}} \times  \max_m \log_2 \Pud \left(1+\gamma^d_{\rm A} \mathbf{h}_{{\rm A}{\rm E}_m}^{d \top} \mathbf{p}_{\rm A}^d \left(\mathbf{h}_{{\rm A}{\rm E}_m}^{d \top}\mathbf{p}_{\rm A}^d \right)^* \right) > \mathcal{R}_{\rm th} \bigg\}
\end{split}
\end{equation}
\normalsize
where $\mathcal{R}_{\rm th} = \frac{N_{\rm data}}{N+N_{\rm cp}}\log_2 \left(1+\gamma^d_{\rm A} \frac{N_{\rm A}}{\Gamma_{\rm AB}}\right) - \mathcal{R}_{\rm data}$. Note that $\mathbf{h}_{{\rm A}{\rm E}_m}^d \mathbf{p}_{\rm A}^d$ is a complex Gaussian random variable with zero mean and unit variance since the precoder, $\mathbf{p}_{\rm A}^d$, has a unit norm. Hence, $\mathbf{h}_{{\rm B}{\rm E}_m}^d \mathbf{p}_{\rm B}^d \left(\mathbf{h}_{{\rm B}{\rm E}_m}^d\mathbf{p}_{\rm B}^d\right)^*$ follows a Chi-square distribution with two degrees of freedom which is basically an exponential distribution with mean equal to one. The term, $1+\gamma^d_{\rm A} \mathbf{h}_{{\rm A}{\rm E}_m}^d \mathbf{p}_{\rm A}^d \left(\mathbf{h}_{{\rm A}{\rm E}_m}^d\mathbf{p}_{\rm A}^d \right)^*$ is therefore a shifted exponential distribution with unit shift and mean $=\gamma_{\rm A}^d$. The probability density function (PDF) of the product of shifted exponential distributions, $ \\ y =  \Pud \left(1+\gamma^k_{\rm A} \mathbf{h}_{{\rm A}{\rm E}_m}^d \mathbf{p}_{\rm A}^d \left(\mathbf{h}_{{\rm A}{\rm E}_m}^d\mathbf{p}_{\rm A}^d\right)^* \right) $ can be expressed as \cite{yilmaz2009product} 
\begin{equation}
\begin{split}
\small
p_{ Y}(y) =\left(\! \Pud \! \frac{1}{\gamma^d_{\rm A}} \exp\left(\frac{1}{\gamma^d_{\rm A}}\right) \right) \times \mathcal{H}_{0,N_{\rm data}}^{N_{\rm data},0} \Bigg[\frac{y}{\Pud\gamma^d_{\rm A}}   \;\vline\; \begin{matrix}{ (-,-,-)}\\{\left(0,1,\frac{1}{\gamma^1_{\rm A}}\right), \dots, \left(0,1,\frac{1}{\gamma^{N_{\rm data}}_{\rm A}} \right) }\end{matrix} \Bigg]
\end{split}
\end{equation}
where $\mathcal{H}_{}$ is the generalized upper incomplete Fox's H function \cite{yilmaz2009product}. The notation $(-,-,-)$ means that these coefficients will be eliminated when substituted in the $\mathcal{H}_{}$ function. Considering a single Eve and mutually independent sub-channels, the pdf of Alice-Eve's rate, $\Sud \!\!\! R_{{\rm A}{\rm E}_m}^d$ can be then expressed as
\begin{equation}
\small
\label{abab}
\begin{split}
    &p_{ R_{\rm AE}}(r) = \frac{1}{\log_2 \left( \exp(\kappa^{-1}) \right)} \times \mathcal{H}_{0,N_{\rm data}}^{N_{\rm data},0} \Bigg[2^r\!\!\Pud\!\!\left(\frac{1}{\gamma^d_{\rm A}}\right)\;\vline\; \begin{matrix}{(-,-,-)}\\{\left(1,1,\frac{1}{\gamma^1_{\rm A}}\right), \dots, \left(1,1,\frac{1}{\gamma^{\rm N_{\rm data}}_{\rm A}}\right) }\end{matrix} \Bigg]
    \end{split}
\end{equation}
where $\kappa =  \left(\! \Pud \! \exp^{\frac{1}{\gamma^d_{\rm A}}}\right)$. The $A-B$ link SOP can be expressed as
\begin{equation}
\small
\label{ABNSOPold}
\begin{split}
    P_{\rm AB}^{\rm SOP} &= \Pr \left\{0 \leq \Sud  {R_{\rm s}}_{\rm AB}^d < \mathcal{R}_{\rm data} \right\} \approx \Pr \left\{\Sud R_{\rm AE}^d > \mathcal{R}_{\rm th} \right\} = \int_{\mathcal{R}_{\rm th}}^{\infty}  p_{ R_{\rm AE}}(r) \: dr 
\end{split}
\end{equation}
Substituting with \eqref{abab} in \eqref{ABNSOPold} and integrating, we get the SOP expression in \eqref{ABNSOP} at the top of the next page.
\begin{table*}
\begin{equation}
\begin{split}
\label{ABNSOP}
    P_{\rm AB}^{\rm SOP} = \kappa \times \mathcal{H}_{0,N_{\rm data}+1}^{N_{\rm data}+1,0} \left[2^{\mathcal{R}_{\rm th}}\!\!\Pud\!\!\left(\frac{1}{\gamma^d_{\rm A}}\right)\;\vline\; \begin{matrix}{\bigg(1,1,0\bigg)}\\{\bigg(0,1,0\bigg),\left(1,1,\frac{1}{\gamma^1_{\rm A}}\right), \dots, \left(1,1,\frac{1}{\gamma^{N_{\rm data}}_{\rm A}}\right) }\end{matrix} \right]
\\ \\ \hline
\end{split}
\end{equation}
\vspace{-0.6 cm}
\end{table*}

\subsection{Bob-Alice link SOP using wiretap coding}
Bob transmits secret key packets to Alice over the $\mathcal{U}_{\rm key}$ sub-channels using wiretap coding. When Bob is given the priority in sub-channels selection, he selects the $\mathcal{U}_{\rm key}$ sub-channels that maximize his secrecy rate as shown in Eqn. \eqref{BobSelects1}. Designing the precoder at Bob is similar to Alice with the respective channel parameters, $\mathbf{p}_{\rm B}^k=\frac{(\mathbf{h}_{BA}^k)^*}{|\mathbf{h}_{BA}^k|}$. Hence, the sub-channels assigned to Bob are given by
\begin{equation} 
\label{BobSelects2}
\begin{split}
   \mathcal{U}_{\rm key}(i) &= {\arg} { \max_{k}} \; \left\{ \log_2 \left(1+\gamma^k_{\rm B} \frac{\|\mathbf{h}_{BA}^k\|^2 }{\Gamma_{\rm BA}}\right) \right.  - \left. \max_{m} \log_2 \left(1+\gamma^k_{\rm B} \|\mathbf{h}_{BE_m}^k\|^2 \right) \right\}, \\   &\forall i \in \{1,2,.., N_{\rm key}\}, k \neq \left\{\mathcal{U}_{\rm key}(1),.., \mathcal{U}_{\rm key}(i-1)\right\}
   \end{split}
\end{equation}
\normalsize
At high SNR, Eqn. \eqref{BobSelects2} is approximated as follows
\begin{equation}
\small
\label{BobSelects3}
   \begin{split}
   \mathcal{U}_{\rm key}(i) &= {\arg} { \max_{k}} \; \left\{ \frac{\|\mathbf{h}_{BA}^k\|^2 }{ \max_{m} \| \mathbf{h}_{BE_m}^k\|^2} \right\}, \forall i \in \{1,2,.., N_{\rm key}\}, k \neq \left\{\mathcal{U}_{\rm key}(1),.., \mathcal{U}_{\rm key}(i-1)\right\}
   \end{split}
\end{equation}
\normalsize
When the number of Bob's transmit antennas is large, $\|\mathbf{h}_{BA}^k\|^2 \approx N_{\rm B}$ and the $B-{\rm A}$ link rate is given by
\begin{equation}\label{rateapprox}
\begin{split}
R_{\rm BA}^k \approx \frac{N}{N+N_{\rm cp}}\log_2 \left(1+\gamma^k_{\rm A} \frac{N_{\rm B}}{\Gamma_{\rm BA}}\right)
\end{split}
\end{equation}
Following the same steps to write down the achievable secrecy rates for the ${\rm B}-{\rm A}$ link secret-key transmissions, the secret keys instantaneous secrecy rate is given by
\begin{equation}\label{eq:BAsecRate}
\begin{split}
\sum_{k \in \mathcal{U}_{\rm key}} & {R_{\rm s}}_{\rm BA}^k = \min_m \sum_{k \in \mathcal{U}_{\rm key}}\left[R_{\rm BA}^k- R_{{\rm B}{\rm E}_m}^k\right]^+ \\
& = \frac{1}{N+N_{\rm cp}}\Biggr[\sum_{k \in \mathcal{U}_{\rm key}} \log_2 \left(1+\gamma^k_{\rm A} \frac{N_{\rm B}}{\Gamma_{\rm BA}}\right)  - \max_m \sum_{k \in \mathcal{U}_{\rm key}}  \log_2 \left(1+\gamma^k_{\rm A} \mathbf{h}_{{\rm B}{\rm E}_m}^{k \top} \mathbf{p}_{\rm B}^k \left(\mathbf{h}_{{\rm B}{\rm E}_m}^{k \top}\mathbf{p}_{\rm B}^k\right)^* \right)\Biggr]^+
\end{split}
\end{equation}
where $\mathbf{h}_{{\rm B}{\rm E}_m}^k \mathbf{p}_{\rm B}^k$ is a Gaussian random variable with zero mean and unit variance. Hence, $\\ \mathbf{h}_{{\rm B}{\rm E}_m}^k \mathbf{p}_{\rm B}^k \left(\mathbf{h}_{{\rm B}{\rm E}_m}^k\mathbf{p}_{\rm B}^k\right)^*$ follows a Chi-square distribution with two degrees of freedom.
To ensure that Bob's transmissions are secured, the following condition should be satisfied
\begin{equation}\label{RbaCon}
\begin{split}
&\sum_{k \in \mathcal{U}_{\rm key}} {R_{\rm s}}_{\rm BA}^k \ge \mathcal{R}_{\rm key}
\end{split}
\end{equation}
The SOP can be expressed similar to the Alice transmissions SOP wiretap coding in \eqref{ABNSOP} by replacing the parameters with those associated with the Bob-Alice link. However, unlike Alice, Bob knows the instantaneous secrecy rate of his transmissions since he is the intended receiver of all users (i.e., Alice and Eves). Hence, Bob can determine whether or not a given transmission (from him) is perfectly secured by ensuring that the instantaneous secrecy rate condition in \eqref{RbaCon} is satisfied before transmitting any keys.

\section{Secure throughput using secret keys}
If the secret-key queue accumulates keys equivalent to a single data packet length, data packet transmissions can be perfectly secured by OTP encryption. Otherwise, Alice has to employ wiretap coding to secure her transmissions. In both cases, Alice is assigned $N_{\rm data}$ sub-channels if $B-A$ link is not in secrecy outage. Otherwise, Alice is assigned $N$ sub-channels.  Hence, the secure throughput of the system in bits/channel use is given by
\begin{equation}\label{gamma_24444tt}
\small
\begin{split}
\mathcal{T}_{\rm s} = & \Bigg[ \Pr\{Q_{\rm key} (t) \ge {{K}}\} 
 \times \bigg\{  \left(1 - P_{\rm AB}^{\rm OP}(N_{\rm data})  \right) \left(1 - P_{\rm BA}^{\rm SOP} (N_{\rm key})  \right)  +  \left(1 - P_{\rm AB}^{\rm OP}(N)  \right)  P_{\rm BA}^{\rm SOP} (N_{\rm key})   \bigg\}
\\&+ \Pr\{Q_{\rm key} (t) < {{K}}\} \times \bigg\{ \left(1 - P_{\rm AB}^{\rm SOP}(N_{\rm data})  \right) \left(1 - P_{\rm BA}^{\rm SOP} (N_{\rm key})  \right)  +  \left(1 - P_{\rm AB}^{\rm SOP}(N)  \right)  P_{\rm BA}^{\rm SOP} (N_{\rm key})   \bigg\}\Bigg] R_{\rm data}
\end{split}
\end{equation}
where $P^{\rm OP}_{\rm m1-m2}(\beta)$, $P^{\rm SOP}_{\rm m1-m2}(\beta)$ denote respectively, the connection outage probability and the SOP of the $m_{\rm 1}-m_{\rm 2}$ link when $\beta$ sub-channels are used. Note that $P_{\rm AB}^{\rm OP}(\beta) \le P_{\rm AB}^{\rm SOP}(\beta)$ where $P_{\rm AB}^{\rm OP}(\beta)$ is obtained from \eqref{connOutage} while  $P_{\rm AB}^{\rm SOP}(\beta)$ is calculated using \eqref{ABNSOP}. As can be understood from the expression in \eqref{gamma_24444tt}, increasing $ \Pr\{Q_{\rm key} (t)\ge {{K}}\}$ can achieve the perfect security of the system where the secure throughput is determined by the reliability of the Alice-Bob link and completely independent of Eve's presence.

Let $\Pr\{Q_{\rm key} (t)=\ell\}$ denote the probability that Alice's secret-key queue has accumulated $\ell$ key packets at time $t$. Using the Markov chain analysis in Appendix \ref{proofMAC}, the steady state probabilities are given by
\begin{equation}
\begin{split}
\label{poklolll}
\Pr\{Q_{\rm key} (t)=0\}&=\frac{(1-\lambda)}{K}, \;\;\;\;  \Pr\{Q_{\rm key} (t)=\ell\}\! =\!\frac{1}{K} \forall \ell\in \Sigma, \\ \Pr\{Q_{\rm key} (t)=K\}\!&=\!\frac{\lambda}{K}, \;\;\;\; \Pr\{Q_{\rm key} (t)=\nu\}\!=\!0  \ \forall \nu >K
\end{split}
\end{equation}
where $\lambda$ is the mean arrival rate of the secret-key queue, $\Sigma=\{1,2,\dots,K-1\}$, $K\le Q_{\max}$ and $Q_{\max}$ is the secret-key buffer size. From the solution of the Markov chain, we observe that as long as $K <  Q_{\max}$, $Q_{\max}$ has no impact on the security. However, increasing $Q_{\max}$ can improve the security performance since it increases the feasible range of the optimization variable, $K$. Substituting with the steady-state probabilities in \eqref{poklolll} into the secure throughput in \eqref{gamma_24444tt}, we get the secure throughput expression in \eqref{jijoo} at the top of the next page, where the mean arrival rate of the secret-key packets can be expressed as
\begin{equation}
\vspace{-0.5 cm}
\begin{split}
    \lambda &=\left(1-P^{\rm SOP}_{\rm BA}(N_{\rm key})\right) \left( 1- P^{\rm OP}_{\rm AB}(N_{\rm data})\right)  + \left(1-P^{\rm SOP}_{\rm BA}  (N)\right) P^{\rm OP}_{\rm AB}(N_{\rm data}) 
    \end{split}
\end{equation}
\vspace{-0.5 cm}

Our ultimate goal is to maximize the secure throughput ($\mathcal{T}_{\rm s}$) in \eqref{jijoo}. The expression $1-P^{\rm SOP}_{\rm BA}(\beta)=\Pr\{{R_{\rm s}}_{\rm BA} \ge \frac{\mathcal{R}_{\rm data}}{K}\}$ is a function of $K$ and monotonically decreasing with it. if one can ensure  $\frac{1-P^{\rm SOP}_{\rm BA}(\beta)}{K}$ is always equal to $1$, we can ensure that all transmissions can be perfectly secured using secret-keys. The numerator of $\frac{1-P^{\rm SOP}_{\rm BA}(\beta)}{K}$ is monotonically decreasing with $K$, while the denominator is monotonically increasing with it. This highlights the importance of selecting $K$ carefully. 

One can further analyze the parameter $K$ as follows. Let ${B}_{\rm data}= \mathcal{R}_{\rm data} W T$ and ${B}_{\rm key}=B_{\rm data}/K=\mathcal{R}_{\rm data} W T/K$ denote the number of bits per data and secret-key packets, respectively. When $K=1$, the secret key packet size is equal to the data packet size and both have a transmission rate of $\mathcal{R}_{\rm data}$ bits/channel use. If $K=2$, the secret key packet size becomes $B_{\rm data}/2$ bits/channel use and becomes easier to achieve. Increasing $K$, increases the secret-key packets mean arrival rate and decreases their SOP. However, the steady-state probabilities in \eqref{poklolll} decrease with $K$. This decreases the probability that the secret-key queue accumulates $K$ secret-key packets which is necessary to secure an entire data packet. This implies that there is an optimal $K$ such that the secure throughput is maximized which we optimize in Section \ref{SecAnalysis}.B.

\begin{table*}
\begin{equation}
\label{jijoo}
\begin{split}
\mathcal{T}_{\rm s} = & \Bigg[ \frac{\lambda}{K}  \times \bigg\{  \left(1 - P_{\rm AB}^{\rm OP}(N_{\rm data})  \right) \left(1 - P_{\rm BA}^{\rm SOP} (N_{\rm key})  \right) 
+  \left(1 - P_{\rm AB}^{\rm OP}(N)  \right)  P_{\rm BA}^{\rm SOP} (N_{\rm key})   \bigg\}
\\&+ \left( 1- \frac{\lambda}{k}\right)  \bigg\{ \left(1 - P_{\rm AB}^{\rm SOP}(N_{\rm data})  \right) \left(1 - P_{\rm BA}^{\rm SOP} (N_{\rm key})  \right) 
+  \left(1 - P_{\rm AB}^{\rm SOP}(N)  \right)  P_{\rm BA}^{\rm SOP} (N_{\rm key})   \bigg\}\Bigg] R_{\rm data}
\\ \\ \hline
\end{split}
\end{equation}
\vspace{-1 cm}
\end{table*}

\section{Analysis and Insights for the proposed scheme}
\label{SecAnalysis}
\subsection{Illustrative numerical example when $N_{\rm key}= 1$}

Assuming independent sub-channels, the sum over sub-channels for the rates can be approximated as a Gaussian random variable with mean $\mu_\circ=N \mu$ and variance $\sigma_\circ^2=N \sigma^2$. When $N_{\rm A}=N_{\rm B}=1$, using the gap approximation, the instantaneous secrecy rate of the secret key packet transmission over sub-channel $k \in \mathcal{U}_{\rm key}$ is given by
\begin{equation}
\small
\begin{split}
\sum_{k \in \mathcal{U}_{\rm key}} {R_{\rm s}}_{\rm BA}^k&=\left[\sum_{k \in \mathcal{U}_{\rm key}}R_{\rm BA}^k-\sum_{k \in \mathcal{U}_{\rm key}}R_{{\rm B}{\rm E}_m}^k\right]^+  = \frac{\left[\sum_{k \in \mathcal{U}_{\rm key}} \log_2 \left(\frac{1+\frac{|\mathbf{h}_{\rm BA}^k|^2 \gamma^k_{\rm B}}{ \Gamma_{\rm BA}}}{1+{|\mathbf{h}_{{\rm BE}_m}^k|^2 \gamma^k_{\rm B}}}\right)\right]^+}{N+N_{\rm cp}}
\end{split}
\end{equation}
Assuming a single Eve and that only a single OFDM sub-channel is assigned to key transmission ($N_{\rm key} = 1$), the SOP of the key transmission is given by \cite{barros2006secrecy}
\begin{equation}
\label{SOPApprox}
\begin{split}
P_{\rm BA}^{\rm SOP}=1- \frac{1}{1+2^{\mathcal{R}_{\rm key}(N+N_{\rm cp})}} \exp\left(-\frac{2^{\mathcal{R}_{\rm key}(N+N_{\rm cp})}-1}{\gamma_{\rm B}^k/\Gamma^k_{\rm BA}}\right)
\end{split}
\end{equation}
Assuming moderate-to-high SNR levels, we can approximate the secret key transmission SOP as
\begin{equation}
\begin{split}
P_{\rm BA}^{\rm SOP}\approx 1- \frac{1}{1+2^{\mathcal{R}_{\rm key}(N+N_{\rm cp})}} \approx 1-2^{-\mathcal{R}_{\rm key} N}
\end{split}
\end{equation}
where the last approximation holds since we ignored the CP length, $N_{\rm cp}$, relative to number of OFDM sub-channels, $N$. If  $\mathcal{R}_{\rm key}=2$ bits/channel use, the SOP is almost $1$ regardless of the used transmit power since we have a factor of $2^{N+N_{\rm cp}}$ in the denominator. In other words, the SOP is approximately given by
\begin{equation}
\begin{split}
P_{\rm BA}^{\rm SOP}\approx 1- 2^{-2N}
\end{split}
\end{equation}
which is very close to $1$ since $N$ is usually greater than $32$. This is mainly because we assumed only a single sub-channel assigned for key sharing ($N_{\rm key} = 1$) in a single-input single-output (SISO) system.  Hence, we are motivated to optimize the sub-channel assignment between the data and the secret key symbols.


\subsection{Offline parameters optimization} 
Our problem is concerned with the secure throughput which is given by
\begin{equation} 
 \underset{\substack{{{N_{\rm data}} \in\{1,2,\dots,N\}}\\{K \in\{1,2,\dots,Q_{\max}\}}}}{\max:} \  \mathcal{T}_{\rm s}
\end{equation}
Once $N_{\rm data}$ and $K$ are optimized, we get $N_{\rm key} = N - N_{\rm data} $ and $\mathcal{R}_{\rm key} = \mathcal{R}_{\rm data}/K$. Since both optimization parameters are integers, a simple approach to obtain the optimal solution that maximizes the secure throughput is to evaluate the objective function $N \times Q_{\max}$ times. The optimal solution is the one that yields the highest objective function in \eqref{jijoo}. Our proposed approach is summarized in Algorithm \ref{euclid}. Although these computations are performed infrequently and offline, their complexity can be reduced by using alternative computationally-efficient methods such as the interior-point method \cite[Chapter 11]{boyed}.
 
 \begin{algorithm}[t]
\fontsize{9}{6}\selectfont
\caption{Optimization Procedure}\label{euclid}
\begin{algorithmic}[1]
\STATE Generate channel realizations using the channel statistics 
\STATE Set $N_{\rm data} = 1$
\WHILE{$N_{\rm data} \ne N$} 
\STATE Compute $N_{\rm key} = N - N_{\rm data}$
\STATE Set $K=1$
\WHILE{ $K \ne Q_{\rm max}$}
\STATE For each channel realization, compute the Alice-Bob link rate from \eqref{rateapproxxx} and  its secrecy rate outage probability from \eqref{ABNSOP}
\STATE Compute the Bob-Alice secrecy rate using \eqref{eq:BAsecRate} 
\STATE Compute the secure throughput, $\mathcal{T}_{\rm s}$ using \eqref{jijoo}
\STATE Set $K= K+1$
\ENDWHILE
\STATE Set $N_{\rm data} = N_{\rm data} +1 $
\ENDWHILE
\STATE Select $K$ and $N_{\rm data}$ that maximize $\mathcal{T}_{\rm s}$ 
\end{algorithmic}
\end{algorithm}
 
 \subsection{Dynamic parameters optimization} 
\begin{figure}
    \centering
\normalcolor
  \includegraphics[width=0.7\columnwidth]{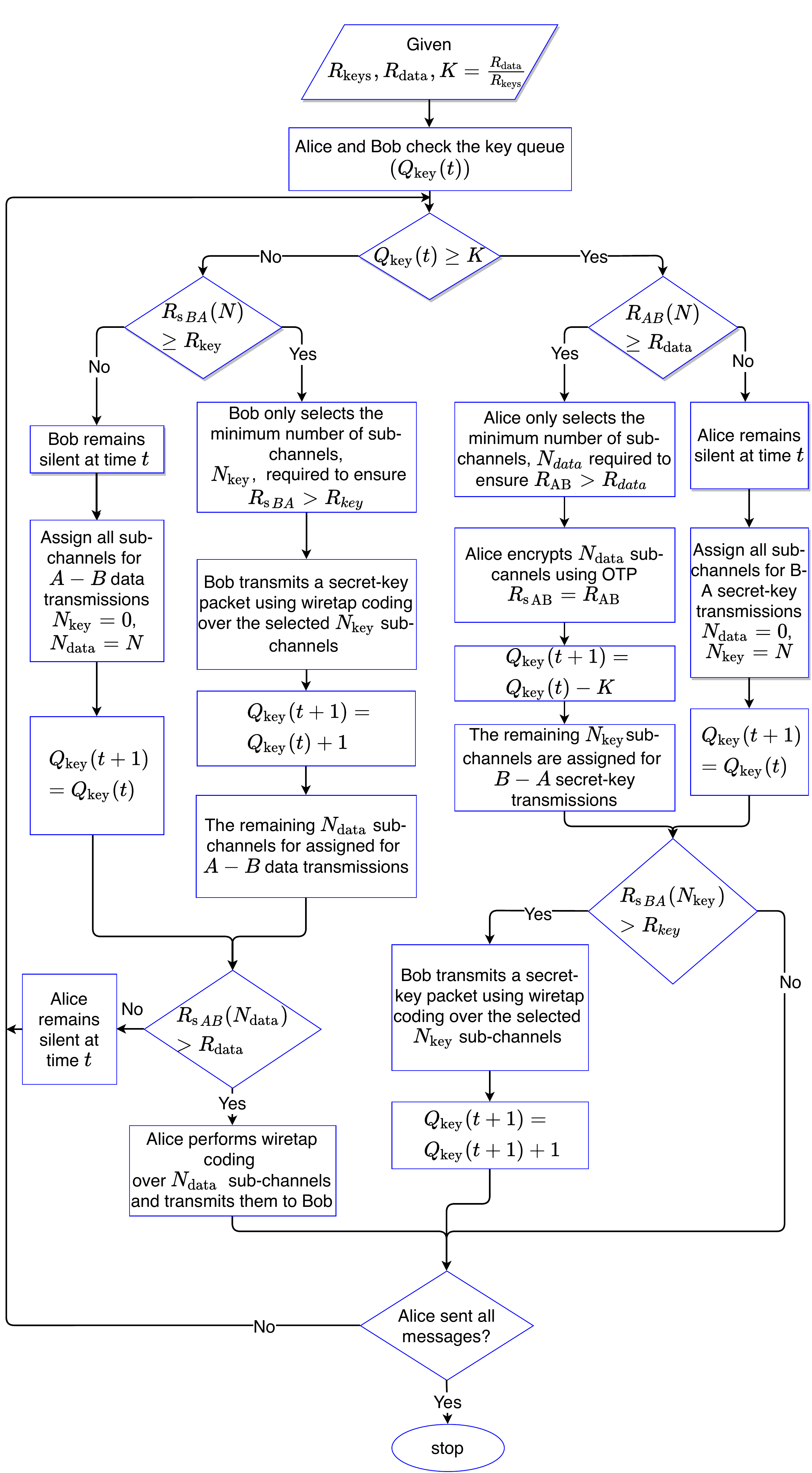}
  \caption{Flow chart of our proposed secret-key-assisted scheme using dynamic sub-channel allocations}\label{fig:flowChart2}
\end{figure}

 The previous optimization suggests using fixed allocation of data and key sub-channels which remains the same as long as the channel statistics remain the unchanged. Another approach to increase the secrecy rate is by relaxing the fixed allocation requirement of the keys and data sub-channels ($N_{\rm key}$ and $N_{\rm data}$). At the intervals of OTP encryption (i.e., the secret-key queue has one data packet length of keys), Alice only selects the minimum number of sub-channels $N_{\rm data}$ with the highest SNR required to ensure that her link is not in connection outage ($R_{\rm AB} \geq \mathcal{R}_{\rm data}$). The remaining sub-channels should be used for ${\rm B}-{\rm A}$ key sharing. Similarly, at the intervals of wiretap coding, Bob, since he knows the instantaneous secrecy rates, uses the minimum number of sub-channels required to ensure his key transmission is not in outage (${R_{\rm s}}_{\rm BA} \geq \mathcal{R}_{\rm key}$). The remaining $N-N_{\rm key}$ OFDM sub-channels are used by Alice to send data. The flow chart of this scheme is shown in Fig. \ref{fig:flowChart2}. 
 

\section{Simulation results}

In this section, we carry out numerical simulations to evaluate the performance of our proposed schemes in terms of the achieved secure throughput. The fading channels are assumed to be complex circularly-symmetric Gaussian random variables with zero mean and unit variances. Two nodes are assumed to try eavesdropping on the ongoing communications, i.e., $M=2$. Unless otherwise stated explicitly, we consider an OFDM system with $N=64$, $N_{\rm cp} = L =8$, $\mathcal{R}_{\rm data}=1.5$ bits/channel use,  ${\gamma^k_{\rm A}} = {\gamma^d_{\rm B}} = 30$ dB, $\Gamma_{\rm AB} = \Gamma_{\rm BA} = 1.2$, and  $Q_{\rm max} = 10$. We assume that the base station, Bob, is equipped with eight transmit antennas while the legitimate user, Alice, is equipped with two transmit antennas. To simplify the power allocation, in all transmissions, we consider an equal-power allocation strategy among all assigned sub-channels. We consider $K=1$ and assume that the fixed key-sharing uses algorithm \ref{euclid} to optimize $N_{\rm data}$ only.

\begin{figure}
    \centering
    \begin{minipage}{0.5\textwidth}
        \centering
        \includegraphics[width=8 cm]{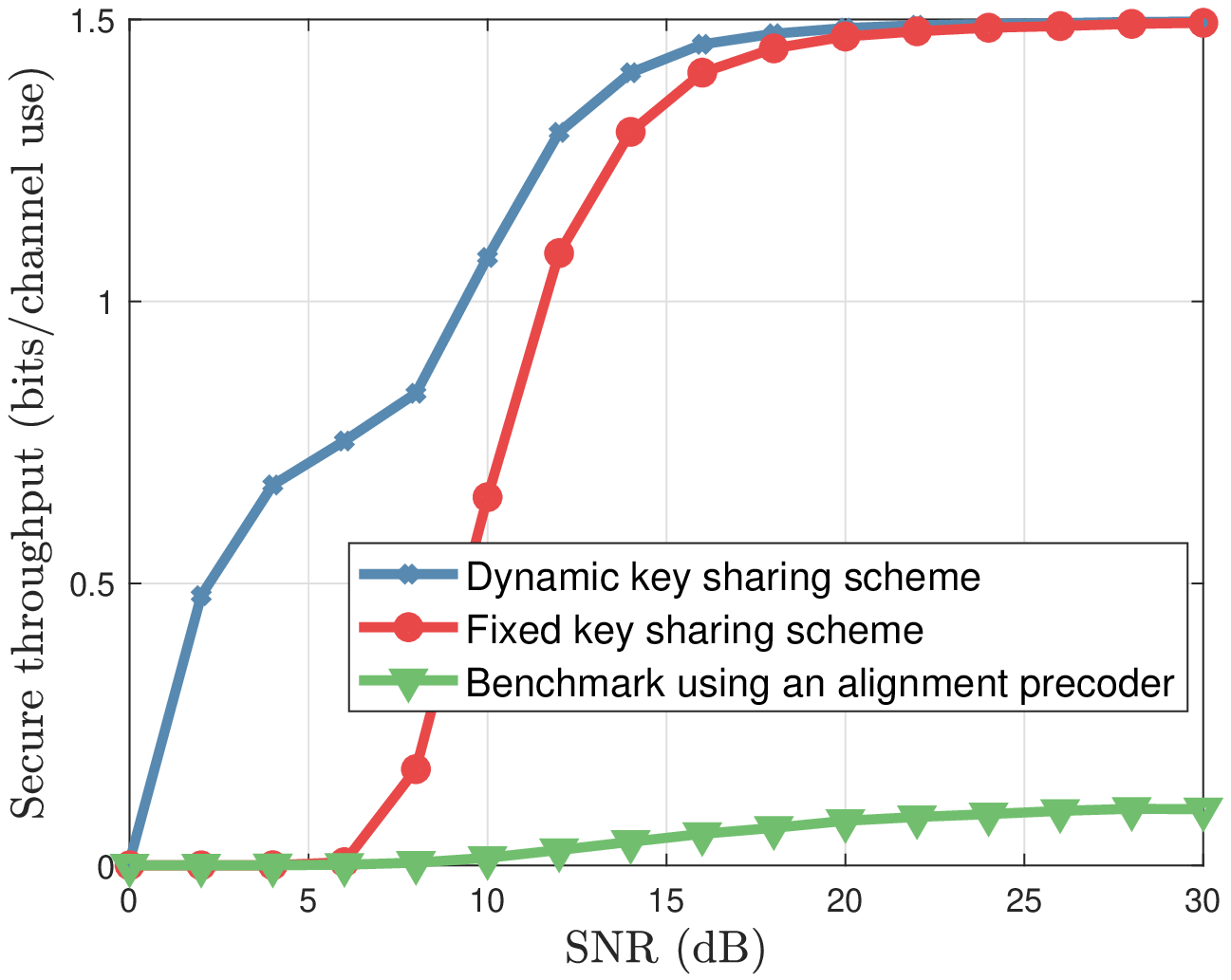} 
                \vspace{-0.6 cm}
  \caption{Secure throughput versus SNR}\label{VsSNR}
    \end{minipage}\hfill
    \begin{minipage}{0.5\textwidth}
        \centering
        \includegraphics[width=8 cm]{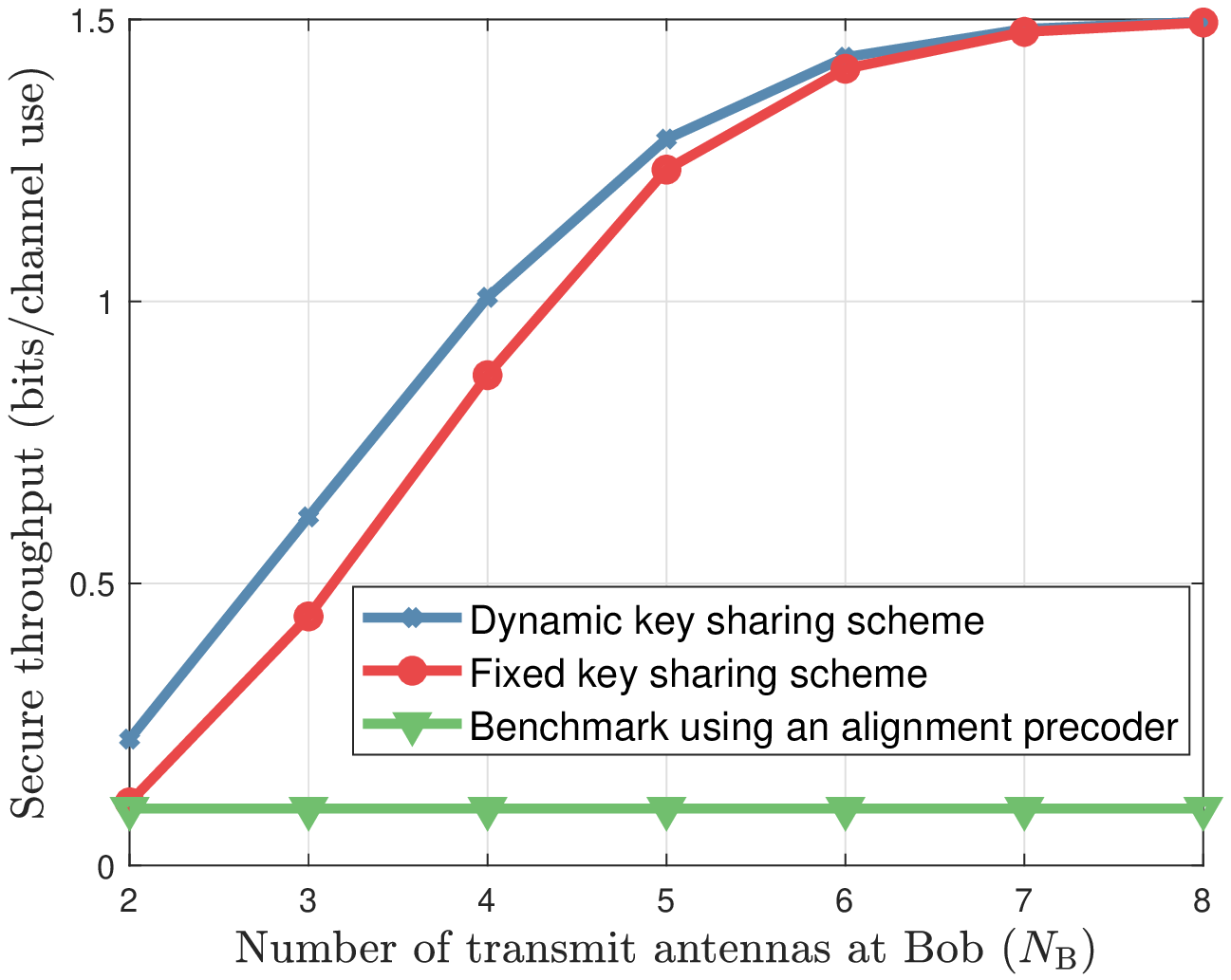} 
        \vspace{-0.6 cm}
  \caption{Secure throughput versus $N_{\rm B}$}
  \label{VsNb}  
    \end{minipage}
        \vspace{-1 cm}
\end{figure}

In Fig. \ref{VsSNR}, we plot the secure throughput performance of our proposed secret-key-assisted schemes relative to the alignment precoding benchmark, where Alice uses all sub-channels for data transmissions and aligns her precoder on each of the Alice-Bob sub-channels. As expected, the secure throughput is monotonically non-decreasing for all schemes. Our schemes achieved the upper bound performance where the achievable secure throughput is equal to the target secrecy rate $\mathcal{R}_{\rm data} = 1.5$ bits/channel use. Although our proposed schemes consume some sub-channel resources for key-sharing, they achieved almost $9$ times gains relative to the benchmark as the OTP encryption scheme protects the data packets from eavesdropping. Our proposed dynamic secret-key sharing scheme achieves the best performance due to its efficient sub-channels allocation between secret-key symbols and data symbols. However, the fixed secret-key sharing scheme introduces a simpler and assumes offline resource allocation with relatively low reduction in performance.

Fig. \ref{VsNb} shows the secure throughput performance versus the number of transmit antennas at the base station, Bob, at $\mathcal{R}_{\rm data} = 1.5$ bits/channel use. The benchmark does not rely on any transmissions from the base station. Hence, its performance remains constant as we increase the number of base station transmit antennas. However, increasing $N_{\rm B}$, enhances Bob's achievable rate for sharing secret keys with Alice which, in turn, enhances our scheme's secure throughput. Increasing the number of transmit antennas at the base station from $2$ to $8$ enhances the performance of our proposed scheme by more than $5$ times while the benchmark remains unchanged.

\begin{figure}
    \centering
    \begin{minipage}{0.5\textwidth}
        \centering
        \includegraphics[width=8 cm]{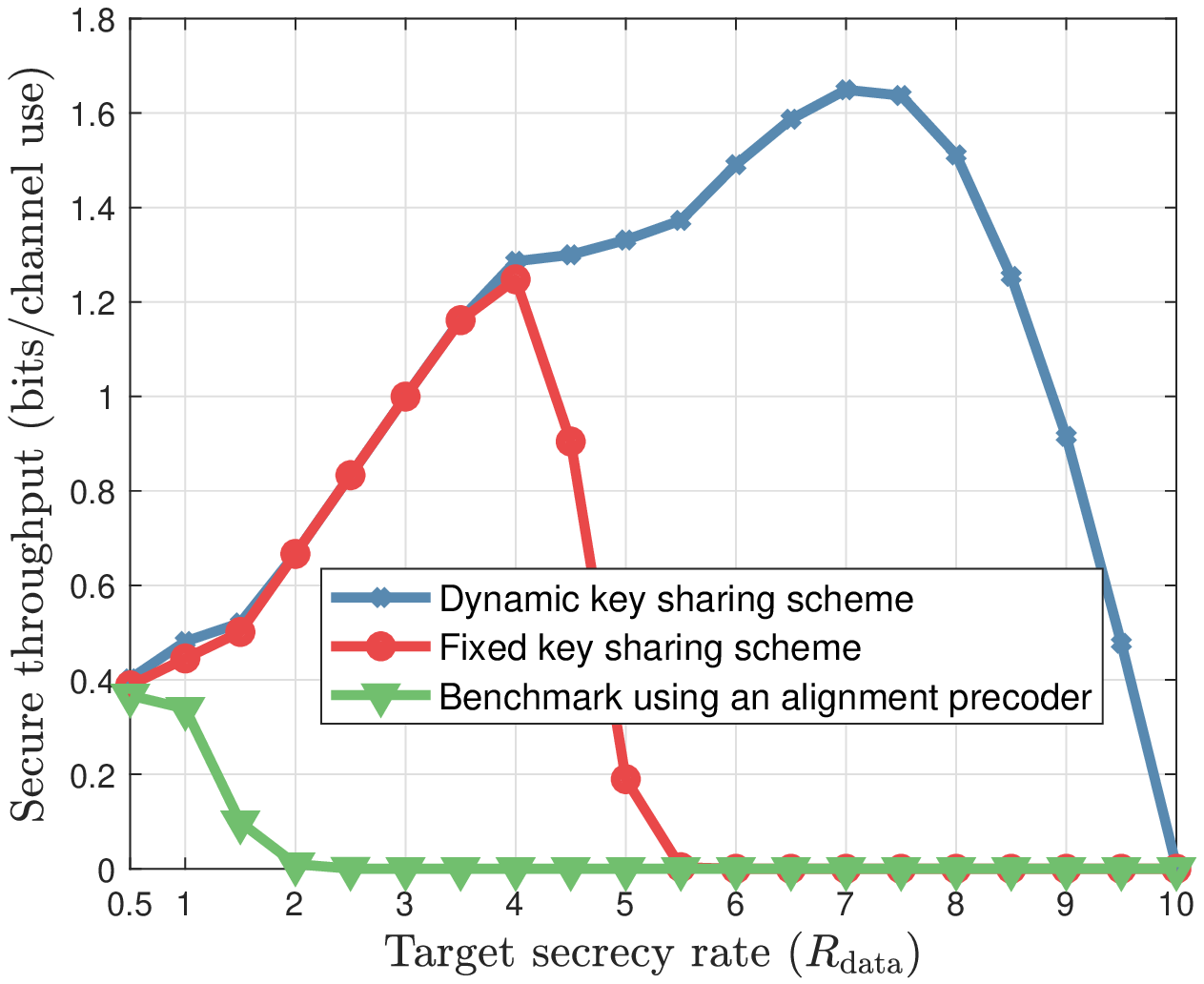} 
                \vspace{-0.6 cm}
  \caption{Secure throughput versus $\mathcal{R}_{\rm data}$}
  \label{VsRdata}
    \end{minipage}\hfill
    \begin{minipage}{0.5\textwidth}
        \centering
        \includegraphics[width=8 cm]{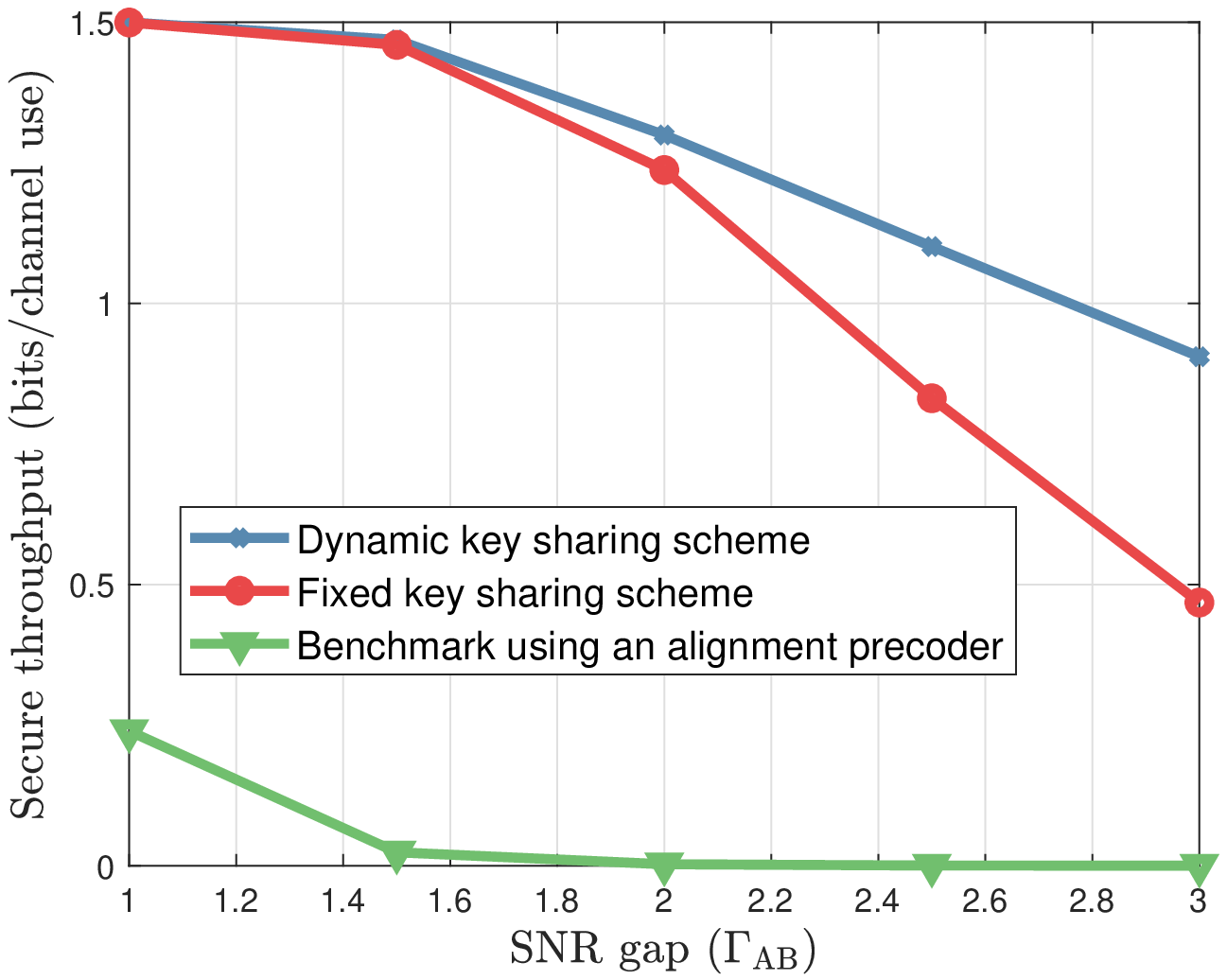} 
        \vspace{-0.6 cm}
  \caption{Secure throughput versus $A-B$ link SNR gap ($\Gamma_{\rm AB}$)}
  \label{VsGammaAB}
    \end{minipage}
        \vspace{-1 cm}
\end{figure}

In Fig. \ref{VsRdata}, we plot the achieved average secure throughput versus the data target secrecy rate, $\mathcal{R}_{\rm data}$, when $K=3$. In all scenarios, the secure throughput keeps increasing until the optimal target secrecy rate is achieved, then, it decreases again. For the benchmark scheme, the optimal value of $\mathcal{R}_{\rm data}$ is $0.5$ bits/channel use while the proposed fixed and dynamic secret-key sharing schemes achieve optimal $\mathcal{R}_{\rm data}$ of $4$ and $7$ bits/channel use, respectively. This demonstrates the significant gains introduced by our proposed schemes.

Fig. \ref{VsGammaAB} depicts the achieved secure throughput versus the SNR gap of the Alice-Bob link. As we increase the SNR gap (harsher target BER requirements), the system becomes less secure for all three schemes. However, our proposed secret-key sharing schemes achieve significantly higher secure throughput performance even at harsh target BER requirements.

\begin{figure}
    \centering
    \begin{minipage}{0.5\textwidth}
        \centering
        \includegraphics[width=8 cm]{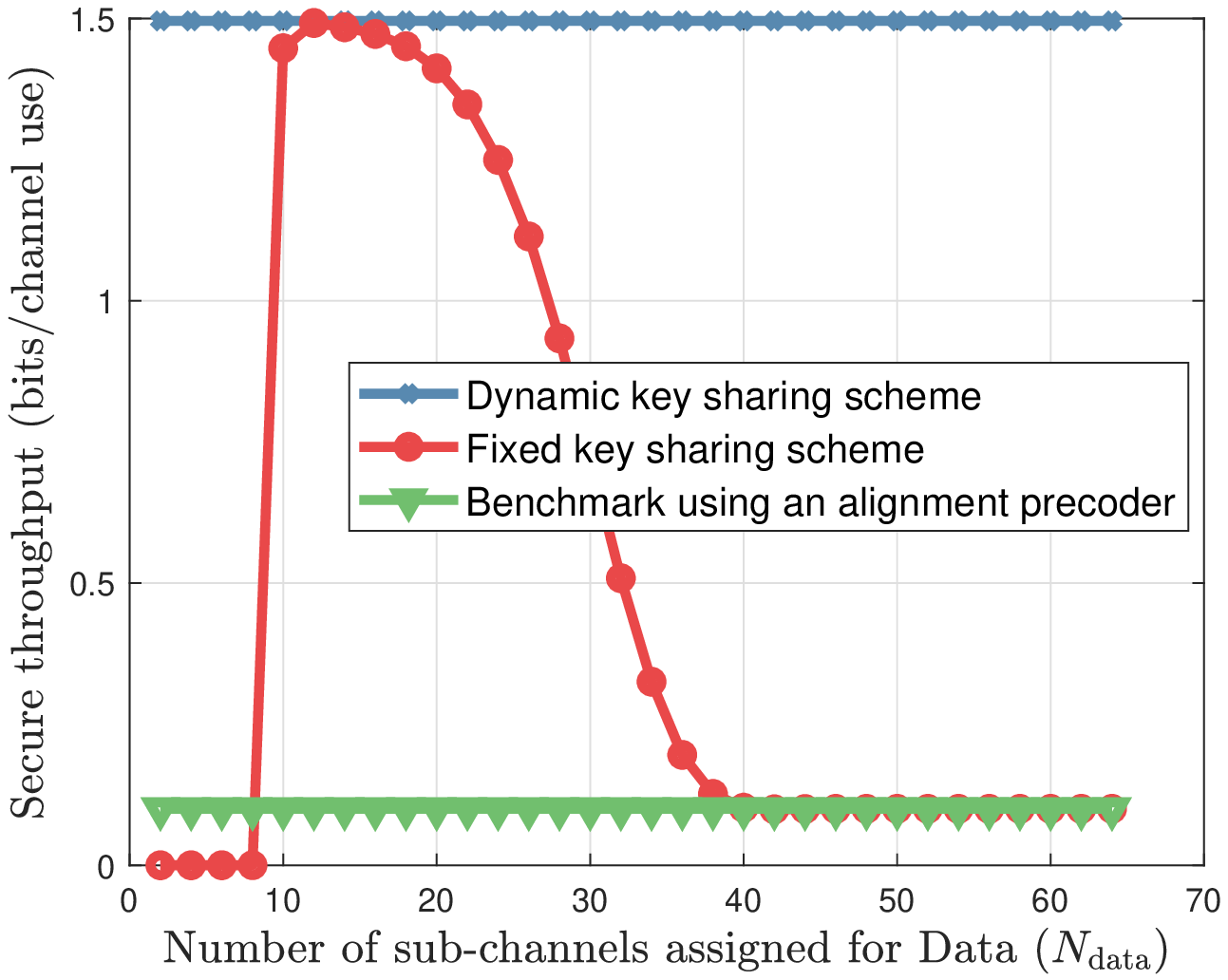} 
                \vspace{-0.6 cm}
  \caption{Secure throughput versus $N_{\rm data}$}
  \label{VsNdata}
    \end{minipage}\hfill
    \begin{minipage}{0.5\textwidth}
        \centering
        \includegraphics[width=8 cm]{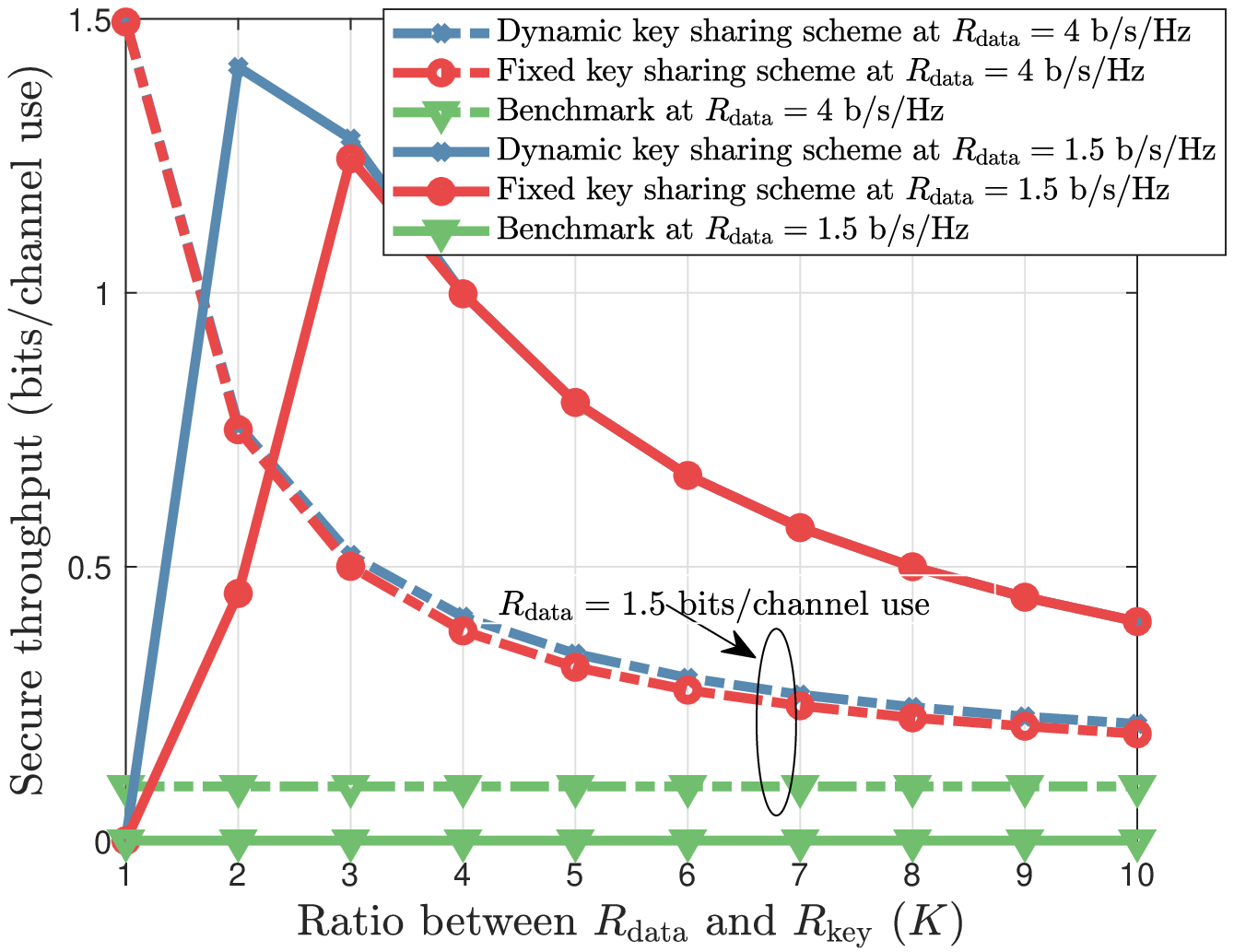} 
        \vspace{-0.6 cm}
  \caption{Secure throughput versus $K$}
  \label{VsKall}
    \end{minipage}
    \vspace{-1 cm}
\end{figure}

Fig. \ref{VsNdata} depicts the achieved average secure throughput versus $N_{\rm data}$ for our proposed key sharing scheme using a fixed sub-channel allocation. For small $N_{\rm data}$, Alice is not assigned sufficient sub-channels to transmit data symbols reliably to Bob. However, as $N_{\rm data}$ increases, significant secure throughput gains are introduced until $N_{\rm data}$ reaches $11$. Then, the secure throughput decreases and saturates at the benchmark secure throughput. High $N_{\rm data}$ (i.e., low $N_{\rm keys}$) restricts Bob's ability in sharing secret keys with Alice which reduces the secure throughput. Note that the ${\rm B}-{\rm A}$ transmissions rely on wiretap coding which requires more sub-channels than OTP encryption. This explains why the optimal $N_{\rm data}$ is $11$ (i.e., optimal $N_{\rm key}$ of $53$).

Fig. \ref{VsKall} depicts the achieved average secure throughput versus the parameter $K$ which represents the ratio between $\mathcal{R}_{\rm data}$ and  $\mathcal{R}_{\rm key}$. We observe an optimal $K$ of one when $\mathcal{R}_{\rm data} = 1.5$ bits/channel use while the optimal $K$ is two and three for the dynamic and fixed key sharing schemes, respectively when $\mathcal{R}_{\rm data} = 4$ bits/channel use. At relatively low and medium target data rates, $\mathcal{R}_{\rm data}$, the optimal value of $K$ is one, which in turn maximizes $\mathcal{R}_{\rm key}$ (i.e. $\mathcal{R}_{\rm key} = \mathcal{R}_{\rm data})$ to ensure maximum key sharing from Bob to Alice. However, at high $\mathcal{R}_{\rm data}$, we have to increase $K$ to ensure the target key rate is reduced $\mathcal{R}_{\rm key} = \mathcal{R}_{\rm data}/K$ and becomes achievable by the wiretap coding schemes. Notice that the OTP encyption-based data transmissions can achieve higher data rates than the key transmissions which rely on wiretap coding. This explains why a relatively low $\mathcal{R}_{\rm key}$ is needed when a high $\mathcal{R}_{\rm data}$ is considered.  

\section{conclusions}
We proposed a new secret-key-assisted scheme to secure uplink MISO transmissions from eavesdropping attacks. Our proposed schemes exploited the sub-channels orthogonality of OFDM transmissions to simultaneously transmit data and secret-key symbols in two opposite directions. The base station, Bob, transmits secret key symbols to Alice and secures his transmissions from eavesdroppers using wiretap coding. Simultaneously, Alice encrypts her data symbols using OTP if she accumulates an equivalent number of key symbols. Otherwise, Alice performs wiretap coding to secure her data transmissions to Bob. We proposed fixed and dynamic sub-channel allocation schemes to divide the sub-channels between data transmissions and secret key sharing. The fixed sub-channel allocation scheme achieves performance close to the dynamic scheme with lower overhead and complexity. We optimized the key and  target data rates and we showed that the data packet length should be relatively large compared to the key packet length to enhance the system's secure throughput. We derived the SOP and the secure throughput expressions for our proposed scheme. Our proposed scheme introduced significant secure throughput gains. For example, at a high SNR of $30$ dB, our proposed scheme achieved more than nine times secure throughput gains relative to the benchmark due to the introduced gains of OTP encryption using secret keys.

 \appendix

\subsection{Wiretap coding}
\label{wiretapCoding}
Consider the case where Alice secures her data packets using wiretap coding.  In a given time slot $t \in  \{1, 2, 3, . . . \}$, Alice adapts her transmission rate $R_{\rm AB}$ to be arbitrarily equal to the link rate to eliminate  the outage events. Let Alice use a codebook $C(2^{nR_{\rm AB}} , 2^{n \mathcal{R}_{\rm data}}, n)$ where $n$ denotes the codeword length in channel uses, $\mathcal{R}_{\rm data}$ denotes the target secrecy rate (i.e. packet size in bits/channel use), $2^{n R_{\rm AB}}$ represents the codebook size, and $2^{n \mathcal{R}_{\rm data}}$ is the number of different confidential messages in the codebook. The $2^{n R_{\rm AB}}$ codewords are randomly and uniformly grouped into $2^{n \mathcal{R}_{\rm data}}$ bins. To transmit a message $ w \in \{1, . . . , 2^ {n \mathcal{R}_{\rm data}}\}$, Alice randomly selects a codeword from bin $w$ and transmits it over the wireless channel. Since Alice does not know the instantaneous CSI of the Alice-Eves links.  We assume that Alice sets a fixed value for its intended secrecy rate, $\mathcal{R}_{\rm data}$ (which represents the spectral efficiency of a single packet transmission). The same wiretap coding scheme is applied when Bob shares his secret keys with Alice. However, Bob knows the channel of his links with Eves. Hence,  Bob only shares the secret keys to Alice when ${\rm A}-{\rm B}$ link is not in secrecy outage.

\subsection{Markov chain analysis}
\label{proofMAC}
In this appendix, we analyze the Markov chain secret-key queue. Let $\pi_i$ denote the probability that the queue has $i$ packets, $\lambda$ is the arrival rate to the queue, where it presents the probability that there is an arrived secret packet at the queue and $f$ is the probability that the queue is serving $K$ packets under certain conditions. To derive the steady-state probabilities of the states, we write down and solve the balance equations at all states. From the balance equation around state $0$, we get
 \begin{equation} 
 \label{eq0x}
 \begin{split}
 \pi_K=\pi_0 \frac{\lambda}{(1-\lambda)f}
\end{split}
\end{equation}
 The state balance equation at states $1$, $2$, $K-1$ and $K$ are given respectively, by
 \begin{equation}
 \small
 \label{eq1}
 \begin{split}
\pi_1 \lambda &= \pi_{0} \lambda+\pi_K \lambda f + \pi_{K+1} (1-\lambda) f \\
\pi_2 \lambda &= \pi_{1} \lambda+\pi_{K+1} \lambda f + \pi_{K+2} (1-\lambda) f \\
\pi_{K-1} \lambda &= \pi_{K-2} \lambda+\pi_{2K-2} \lambda f + \pi_{2K-1} (1-\lambda) f \\
\pi_{K} (1- (1-\lambda) \ \overline{f}) &=\pi_{K-1} \lambda + \pi_{2K} \lambda  f + \pi_{2K+1} (1-\lambda) f
\end{split}
\end{equation}
where $\overline{f} = 1- f$. The balance equation around state $K+1$ is given by
 \begin{equation} 
 \label{eq5}
 \begin{split}
\pi_{K+1} (1- (1-\lambda) \ \overline{f}) =\pi_{K} \lambda \overline{f} + \pi_{2K+1} \lambda f + \pi_{2K+2} (1-\lambda) f
\end{split}
\end{equation}
This should continue up to state $Q_{\max}-K+1$. At $Q_{\max}-K$, the balance equation is
 \begin{equation} 
 \label{eq566}
 \begin{split}
\pi_{Q_{\max}-K} (1- (1-\lambda) \ \overline{f})&=\pi_{Q_{\max}-K-1} \lambda \overline{f} + \pi_{Q_{\max}-1} \lambda  f+ \pi_{Q_{\max}} (1-\lambda)f
\end{split}
\end{equation}
At state $Q_{\max}-K+1$, we have
 \begin{equation} 
 \label{eq566}
 \begin{split}
\pi_{Q_{\max}-K+1} (1- (1-\lambda) \ \overline{f}) =\pi_{Q_{\max}-K} \lambda \overline{f} + \pi_{Q_{\max}} \lambda f
\end{split}
\end{equation}
After that state, the balance equation of state $Q_{\max}-K+1$ is given by
 \begin{equation} 
 \label{eq6xx}
 \begin{split}
\pi_{Q_{\max}-K+2} (1- (1-\lambda) \ \overline{f}) = \pi_{Q_{\max}-K+1} \lambda \overline{f}
\end{split}
\end{equation}
This equation is valid until the final state. That is, around state $Q_{\max}-K+3$, we get
 \begin{equation} 
 \label{eq6xxxxxx}
 \begin{split}
\pi_{Q_{\max}-K+3} = \pi_{Q_{\max}-K+2} \lambda= \pi_{Q_{\max}-K+1} (\lambda \overline{f}) ^2
\end{split}
\end{equation}
From the balance equation around the final state, i.e., state ${Q_{\max}}$, we get
 \begin{equation} 
 \label{eq6xxxx}
 \begin{split}
 \pi_{Q_{\max}-1} =\pi_{Q_{\max}} \frac{f}{\lambda \overline{f}}
\end{split}
\end{equation}
Going one state before that
 \begin{equation} 
 \label{eq6fff}
 \begin{split}
\pi_{Q_{\max}-1} (1- (1-\lambda) \ \overline{f}) = \pi_{Q_{\max}-2} (\lambda \overline{f})
\end{split}
\end{equation}
Hence,
 \begin{equation} 
 \label{eq6fffxx}
 \begin{split}
\pi_{Q_{\max}}   &= \pi_{Q_{\max}-2} \frac{(\lambda \overline{f})^2}{f(1- (1-\lambda) \ \overline{f})} \\
\pi_{Q_{\max}}   &= \pi_{Q_{\max}-k} \frac{(\lambda \overline{f})^k}{f(1- (1-\lambda) \ \overline{f})^{(k-1)}}
\end{split}
\end{equation}
where $k\ge K$. Letting $\mathbf{P}_T$ denote the the Markov chain transition probabilities matrix, the steady-state distribution vector $\mathbf{\Pi}=[\pi_0,\pi_1,\dots,\pi_{Q_{\max}}]$ for the
given transitions is obtained from calculating higher powers of
the matrix \cite{gallager2012discrete} as follows
 \begin{equation} 
 \begin{split}
\mathbf{\Pi}=\mathbf{\Pi}^{(0)}\mathbf{P}_T^{\infty}
\end{split}
\end{equation}
since the Markov chain is irreducible with ergodic (i.e., aperiodic
positive-recurrent) states. All initial distributions, $\mathbf{\Pi}^{(0)}$, will eventually lead to the same steady-state \cite{gallager2012discrete}.
\begin{figure}
    \centering
\normalcolor
  \includegraphics[width=0.65\columnwidth]{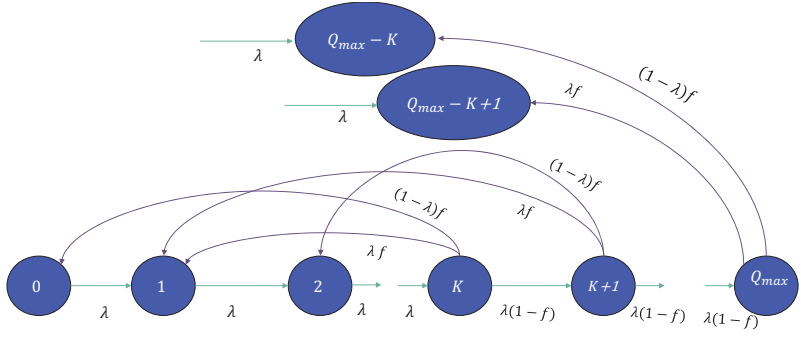}
    \vspace{-0.2 cm}
  \caption{Markov chain model for the energy queues. State-self transitions are omitted from the graph for visual clarity.}\label{figmc1}
  \vspace{-0.5 cm}
\end{figure}

\subsection{Approximated Markov Chain Analysis}
\label{proofMAC}
To obtain closed-form expressions of the Markov chain, we use the assumption that the queue absorbs $K$ packets once it has them. That is, writing the balance equations at different states based on that assumption, we get
 \begin{equation} 
 \small
 \label{eq0x}
 \begin{split}
 \pi_K=\pi_0 \frac{\lambda}{(1-\lambda)}
\end{split}
\end{equation}
 The state balance equations are given by
 \begin{equation}
 \small
 \label{eq1}
 \begin{split}
\pi_1 &= \pi_{0} +\pi_K =\pi_{0}+\pi_0 \frac{\lambda}{(1-\lambda)}=\frac{\pi_0}{(1-\lambda)} \\
\pi_{k} &= \pi_{k-1}, \;\;\;\; \forall k \in \{2,\dots, K-1\}
\\
\pi_{K} &=\pi_{K-1} \lambda
\end{split}
\end{equation}
From the last balance equation, we have
 \begin{equation} 
 \begin{split}
 \label{pokoop}
\pi_{K-1} =\pi_{K-2}=\dots=\pi_1=\frac{\pi_{K}}{\lambda}
\end{split}
\end{equation}
Using the identity that the sum over the state probabilities is the unity, we get
 \begin{equation} 
 \small
 \begin{split}
 \label{pokoopxxx}
&\pi_0+\pi_1+\pi_2+\dots+\pi_{K-1}+\pi_K=1\\
& \pi_0+(K-1)\frac{\pi_{K}}{\lambda}+\pi_K=1
\\ & \pi_0 + (\frac{(K-1)}{\lambda}+1) \pi_0 \frac{\lambda}{(1-\lambda)}=1
\end{split}
\end{equation}
Thus,
 \begin{equation} 
 \begin{split}
 \pi_0 =\frac{(1-\lambda)}{K},\;\; \pi_K &=\frac{\lambda}{K}, \;\;  \pi_k =\frac{1}{K} \forall k \ne 0, K
\end{split}
\end{equation}

\ifCLASSOPTIONcaptionsoff
  \newpage
\fi



%
%
\bibliographystyle{IEEEtran}
\bibliography{references_new}

\end{document}